\documentclass{aa}

\usepackage{xcolor}
\usepackage{graphicx}
\usepackage{txfonts}
\usepackage{lipsum}
\usepackage{subcaption}         
\usepackage{lscape}             
\usepackage{placeins}           

\usepackage{hyperref}
\hypersetup{
    colorlinks=true,
    linkcolor=black,
    citecolor=blue,
    }

\begin{document}

   \title{Globular cluster Observations for Luminosity-dispersion Distances (GOLD): Project description and velocity catalogue}

\titlerunning{GOLD: Project description and velocity catalogue}

%
%
%

   \author{K. Fahrion\inst{1}\fnmsep\corrauth{katja.fahrion@univie.ac.at}        
        \and M. A. Beasley\inst{2}\fnmsep\inst{3}\fnmsep\inst{4} 
        \and A. Gvozdenko\inst{5}
        \and I. Ortega-Casas\inst{2}\fnmsep\inst{3}
        \and A. Ferre-Mateu\inst{2}\fnmsep\inst{3}\fnmsep\inst{4}
        \and F. I. Aros\inst{1}
        \and D. Dold\inst{6}
        \and T. Jerabkova\inst{7}
        \and S. S. Larsen\inst{8}
        \and O. M\"uller\inst{9}
        \and M. Rejkuba\inst{10}
        \and K. L. Rhode\inst{11}
        \and G. van de Ven\inst{1}
        }

   \institute{Department of Astrophysics, University of Vienna, T\"{u}rkenschanzstra{\ss}e 17, 1180 Wien, Austria 
   \and Instituto de Astrofísica de Canarias, Calle Vía Láctea, E-38206 La Laguna, Spain 
        \and    
            Departamento de Astrofísica, Universidad de La Laguna, E-38206 La Laguna, Spain 
        \and 
            Centre for Astrophysics and Supercomputing, Swinburne University, John Street, Hawthorn VIC 3122, Australia 
        \and
            Department of Physics, Centre for Extragalactic Astronomy, Durham University, South Road, Durham DH1 3LE, UK 
        \and
            Faculty of Mathematics and Research Network Data Science, University of Vienna, Kolingasse 14-16, 1090 Wien, Austria 
        \and
            Department of Theoretical Physics and Astrophysics, Faculty of Science, Masaryk University, Kotlárská 2, Brno 611 37, Czechia 
        \and
            Department of Astrophysics/IMAPP, P. O. Box 9010, 6500 GL Nijmegen, The Netherlands 
        \and
        Institute of Physics, Laboratory of Astrophysics, Ecole Polytechnique Fédérale de Lausanne (EPFL), 1290 Sauverny, Switzerland 
        \and
        European Southern Observatory, Karl-Schwarzschild Strasse 2, 85748 Garching, Germany 
        \and
            Department of Astronomy, Indiana University, 727 East Third Street, Bloomington, IN 47405 USA 
            }

   \date{\today}

 
  \abstract
   {We introduce and describe the Globular cluster Observations for Luminosity-Dispersion distances (GOLD) project, which has the principal aim of determining independent distances based on the globular cluster (GC) absolute magnitude-velocity dispersion ($M_V - \sigma$) relation calibrated with Milky Way GCs with Gaia distances. The project comprises observations of GCs in 19 early- and late-type galaxies taken with FLAMES/GIRAFFE on ESO's Very Large Telescope. So far, 17 galaxies have been observed. The spectra cover the wavelength range 8483--9000 \AA\,, including the calcium triplet region at a spectral resolution of $R\sim18,000$. The spectra provide line-of-sight velocities with typical uncertainties of $\Delta v = 1-2$ km  s$^{-1}$. Using primarily velocities to separate extragalactic sources from foreground stars, we present here our initial sample of 1062 GCs and 101 ultra-compact dwarf galaxies (UCDs). For more than 650 of these objects we present the first velocity measurement. We discuss the available photometry and size measurements, and show that our velocities are in good agreement with previous studies. Our velocity catalogues are made publicly available.
   Beyond the principal aim of obtaining precision distances, additional science to be explored with these data includes host galaxy halo dynamics and clustering analysis, mass-to light ratios of GCs and UCDs in the sample, the stellar initial mass function and black hole occupation fraction of the GCs, and chemical abundances from high-dispersion spectra.}


   \maketitle
\nolinenumbers 

\section{Introduction}
Globular clusters (GCs) are compact, massive star clusters (effective radius $R_\text{eff} \sim 3$ pc, $M = 10^4 - 10^6 M_\sun$) characterised by old, often metal-poor stellar populations \citep{BrodieStrader2006, Beasley2020}. They are found in almost all galaxies, and while dwarf galaxies often only host a few GCs, the most massive galaxies in the Universe can host tens of thousands \citep{Harris2013, Harris2017}. Due to their old ages, GCs are often regarded as tracers of early star formation and hierarchical galaxy assembly processes (e.g. \citealt{Forbes2018, Massari2026})

In the Milky Way (MW) and its satellites, GCs can be spatially resolved into their constituent stars, enabling detailed studies of their dynamical and chemical properties (e.g. \citealt{Gratton2019, VasilievBaumgardt2021, MiloneMarino2022, Clontz2024}). In contrast, ground- and space-based imaging of many nearby galaxies has provided insights into the overall properties of extragalactic GC systems (e.g. \citealt{Peng2008, Jordan2009, Jordan2015, Cantiello2020}). For example, most massive galaxies ($M_\ast > 10^{10} M_\sun$; \citealt{Peng2006}) host GC systems with bimodal or complex colour distributions that likely reflect the presence of both red, in situ born GCs as well as blue, accreted populations (e.g. \citealt{ZepfAshman1993, Peng2008, Beasley2018, Harris2023}). 

While imaging has confirmed the connection of GCs with both in situ and accreted galaxy components, additional spectroscopy is required to trace a galaxy's evolutionary history.
With spectroscopy, line-of-sight (LOS) velocities of individual GCs are derived, which are used to confirm membership in a host galaxy and to perform dynamical modelling (e.g. \citealt{Pota2015, Forbes2017}). In many cases, such modelling finds kinematic differences between the red and blue populations (e.g. \citealt{Schuberth2006, Chaturvedi2022}). Additionally, spectroscopy is required to measure stellar population properties such as the ages and metallicities of extragalactic GCs (e.g. \citealt{Usher2012, Villaume2019, Fahrion2020b, Forbes2022, Fahrion2025b}). 

In general, most spectroscopy of extragalactic GCs has been based on multi-object spectroscopy at low and medium resolution ($R = 1000 - 6000$) at optical or near-infrared wavelengths. For example the SAGES Legacy Unifying Globulars and GalaxieS (SLUGGS) survey \citep{Brodie2014, Forbes2017} used the DEIMOS spectrograph on the Keck II telescope to obtain spectra of GCs around about 25 galaxies at a resolution of $R \sim 5500$. Such spectra have typical velocity uncertainties of $5 - 20$ km s$^{-1}$, depending on the signal-to-noise (e.g. \citealt{Brodie2014}). Furthermore, the spectral resolution is too low to measure internal stellar velocity dispersions of typical GCs, and they might be at the limit of what is needed to determine velocity dispersion in ultra-compact dwarf galaxies (UCDs; e.g. \citealt{Hau2009}). For this reason, our understanding of the internal dynamics of extragalactic GCs is limited to the nearest galaxies. The GCs of M31 and Centaurus\,A (NGC\,5128) have been observed at a higher spectral resolution to determine their internal velocity dispersions and the mass-to-light ratios ($M/L$, \citealt{Strader2011, Dumont2022}), but the lack of large spectroscopic samples at higher resolutions limits our general understanding of the internal dynamics of extragalactic GCs.

Due to the lack of high-resolution spectroscopy, very little is known about individual elemental abundances of extragalactic GCs, even though promising approaches to extracting reliable abundances also exist for low-resolution spectroscopy \citep{CabreraZiri2022}. In contrast, high-resolution integrated-light spectroscopy ($R > 10,000$) is currently mostly limited to GCs of the MW and its nearest neighbours (e.g. \citealt{Sakari2016, Colucci2017}). Such work has already revealed interesting chemical abundance patterns in GCs of Local Group dwarf galaxies that likely trace different star formation conditions and timescales \citep{Hernandez2018, Larsen2022, Gvozdenko2024}, and shows promise for the use of GCs as tracers for extragalactic archaeology (\citealt{RecioBlanco2021, Monty2024, Gvozdenko2026}).

In this paper, we introduce the Globular Cluster Observations for Luminosity-dispersion Distances (GOLD) project, an effort to collect high-resolution spectroscopy ($R = 18,000$) of GC systems in galaxies out to 20 Mpc with the FLAMES instrument at the Very Large Telescope (VLT). The main goal is to measure distances from the relation between internal GC velocity dispersions and magnitudes. Similar to the Faber-Jackson relation of galaxies \citep{FaberJackson1976}, the GC velocity dispersion (GCVD) relation is based on the Virial theorem. While the velocity dispersion traces the total dynamical mass, the absolute magnitude serves as a proxy for the luminous mass. Assuming a common mass-to-light ratio ($M/L$) for GCs, a tight log-linear relation between both quantities emerges that can be used to measure distances if a zeropoint is established. While the $M/L$ value is not strictly constant, a narrow range is expected due to the overall old ages of GCs, lack of dark matter, and their narrow size distribution (peaking around $R_\text{eff} \sim 3$ pc; \citealt{BrodieStrader2006, Masters2010}), and is supported by observations. For example, \cite{Baumgardt2020} find $M/L$ values for Galactic GCs in a narrow range between 1.4 and 2.5 $M_\odot/L_\odot$, with a mean value of 1.83 $M_\odot/L_\odot$ and a standard deviation of 0.24 $M_\sun/L_\sun$ around the mean. In extragalactic systems, dynamical $M/L$ values can also be directly measured if size measurements are available (e.g. \citealt{Rejkuba2007, Strader2009, Dumont2022, Fahrion2026a}). With GOLD, we plan to explicitly test the effect of $M/L$ values on GCVD distances.

While the underlying method was conceptually proposed several decades ago \citep{Paturel1992, Djorgovski1995, Strader2009}, \cite{Beasley2024} recently collected velocity-dispersions and $V$-band magnitudes for GCs in the MW and M31 to calibrate the GCVD relation using Gaia distances for anchoring. \cite{Beasley2024} used literature velocity dispersions to derive distances to Centaurus\,A and Local Group dwarf galaxies with uncertainties as low as $3-4\%$. Given that GCs are found in all massive galaxies regardless of galaxy type or morphology, these encouraging results may imply that the GCVD method can develop into an alternative distance measurement method competitive with established methods such as the tip of the red giant branch (TRGB; e.g. \citealt{Lee1993}), Cepheids (e.g. \citealt{Madore1991}), or surface brightness fluctuations (SBFs; e.g. \citealt{Tonry1988}).

At the moment, applications of the GCVD method are still limited. It has been applied to the ultra-diffuse galaxy NGC\,1502-DF2 (see \citealt{Beasley2025} and \citealt{vanHeumen2026}) and M104 (NGC\,4594, \citealt{Fahrion2026a}) using dedicated FLAMES observations in the calcium triplet (CaT) spectral region. With GOLD, we aim to scale up this effort of measuring distances by collecting the largest sample of high-resolution spectroscopy of GCs beyond the Local Group, which will allow us to thoroughly test the GCVD approach for extragalactic distance measurements.
Additionally, we aim to use the chemistry-sensitive CaT region to extract chemical abundances for a large sample of extragalactic GCs, use precision LOS velocities to model dark matter haloes, and study the internal dynamics of GCs including their dynamical $M/L$ values in different environments.

Besides GCs, GOLD also covers UCDs as a complementary type of compact stellar system \citep{Hilker1999, Drinkwater2000, Liu2015}. They tend to be both brighter and larger (with sizes up to $\sim$100 pc) than typical GCs, and multiple scenarios have been discussed concerning their origin and nature: (1) UCDs might be the high-luminosity tail of the GC luminosity function, (2) the products of mergers of massive star clusters, (3) the remnant nuclei of disrupted galaxies stripped by tidal interactions, or (4) primordial dwarf galaxies (e.g. \citealt{Bekki2001, Mieske2008, Fellhauer2002, Dabringhausen2010, Chilingarian2011}). The tidal stripping origin for at least a sub-population of UCDs is currently the preferred scenario, consistent with the findings of high dynamical-to-stellar mass ratios (e.g \citealt{Hasegan2005, Hilker2007, Rejkuba2007, Mieske2008}) and the discovery of supermassive black holes in a few systems (e.g. \citealt{Seth2014, Ahn2018}). With GOLD, we will not only enlarge the sample of spectroscopically confirmed UCDs, but also provide precision velocity dispersion measurements to study $M/L$ ratios on a population level for both UCDs and GCs.

In this first paper, we introduce the current status of the GOLD project, present the data, and publish our initial GC and UCD LOS velocity catalogue for 1062 GCs and 101 UCDs in 17 galaxies (see App. \ref{app:vel_catalogue}). Velocity dispersions and other science outputs will be published in future works.
Section \ref{sect:galaxy_sample} introduces the galaxy sample and Sect. \ref{sect:data} presents the observing strategy and data reduction procedures. Section \ref{sect:methods} describes the data analysis to obtain LOS velocities and how the preliminary sample was selected. Initial results showcasing the GC sample are presented in Sect. \ref{sect:properties}. We discuss additional science goals for the GOLD project in Sect. \ref{sect:science} and conclude in Sect. \ref{sect:conclusions}.

\begin{table*}[htp]
    \centering
    \caption{Sample of galaxies observed in GOLD.}
    \begin{tabular}{c r r c r r r r r c}\hline\hline
    Galaxy & RA (J2000) & DEC (J2000) & Morphology & $v_\text{helio}$ & Distance & Ref. & Environment & log($M_\ast/M_\odot$) & Ref.  \\
    & (deg) & (deg) & & (km s$^{-1}$) & (Mpc) \\ 
    (1) & (2) & (3) & (4) & (5) & (6) & (7) & (8) & (9) & (10) \\ \hline
M104 & 189.997633 & $-$11.623054 & SA(s)a & 1089 & 9.6 & 1 & Sombrero & 11.32 & 6 \\
M49 & 187.444900 & $+$8.000400 & E2 & 981 & 16.5 & 2 & Virgo & 11.73 & 7 \\
M60 & 190.916545 & $+$11.552691 & E2 & 1110 & 16.3 & 2 & Virgo & 11.53 & 7 \\
M84 & 186.265597 & $+$12.886983 & E1 & 1017 & 17.3 & 2 & Virgo & 11.37 & 7 \\
M86 & 186.549225 & $+$12.945969 & E3 & $-$224 & 16.9 & 2 & Virgo & 11.46 & 7 \\
M87 & 187.705931 & $+$12.391123 & cD pec & 1284 & 16.3 & 2 & Virgo & 11.48 & 7 \\
M89 & 188.915864 & $+$12.556342 & E0-1 & 340 & 15.8 & 2 & Virgo & 10.98 & 7 \\
M105 & 161.956667 & $+$12.581631 & E1 & 907 & 10.4 & 2 & Leo I & 10.76 & 8 \\
NGC\,4636 & 190.707639 & $+$2.687764 & E0-1 & 938 & 16.8 & 2 & Virgo & 10.51 & 9 \\
NGC\,4697 & 192.149576 & $-$5.800690 & E6 & 1241 & 11.6 & 2 & NGC\,4697 & 11.07 & 10 \\
NGC\,1399 & 54.621179 & $-$35.450742 & E1 pec & 1425 & 19.0 & 3 & Fornax & 11.42 & 11 \\
NGC\,1404 & 54.716321 & $-$35.594392 & E1 & 1933 & 18.7 & 3 & Fornax & 11.14 & 11 \\
NGC\,1380 & 54.114958 & $-$34.976300 & SA0 & 1878 & 20.1 & 3 & Fornax & 10.79 & 11 \\
NGC\,253 & 11.888058 & $-$25.288800 & SAB(s)c & 242 & 3.7 & 4 & Sculptor & 10.52 & 12 \\
NGC\,300 & 13.722694 & $-$37.684213 & SA(s)d & 147 & 2.1 & 4 & Sculptor & 9.41 & 13 \\
NGC\,247 & 11.785642 & $-$20.760400 & SAB(s)d & 174 & 3.7 & 4 & Sculptor & 9.50 & 13 \\
NGC\,55 & 3.723342 & $-$39.196628 & SB(s)m? &131 & 2.1 & 4 & Sculptor & 9.48 & 13 \\
NGC\,7793 & 359.457308 & $-$32.591028 & SA(s)d & 224 & 3.6 & 4 & Sculptor & 9.70 & 13 \\
NGC\,3115 & 151.308029 & $-$7.718606 & S0\^- & 686 & 10.2 & 5 & NGC\,3115 & 11.04 & 14 \\
\hline
    \end{tabular}
    \label{tab:galaxy_sample}
    \tablefoot{(1) Galaxy name, (2, 3) sky coordinates, (4) morphology from the NASA/IPAC Extragalactic Database (NED), (5) systemic velocity, (6) distance, (7) distance reference (1 - \cite{McQuinn2016}, 2 - \citealt{Anand2025_Virgo}, 3 - \citealt{Anand2024_Fornax}, 4 - extragalactic distance database \citealt{Anand2021}, 5 - \citealt{Peacock2015}), (8) environment, (9) stellar mass, (10) stellar mass reference (6 - \citealt{Karachentsev2020}, 7 - \citealt{Peng2008}, 8 - \citealt{Revnivtsev2008}, 9 - \citealt{OSullivan2018}, 10 - \citealt{Cappellari2013}, 11 - \citealt{Liu2019}, 12 - \citealt{Bailin2010}, 13 - \citealt{Dooley2017}, 14 - \citealt{Guerou2016}).}
\end{table*}

\begin{figure}
    \centering
    \includegraphics[width=0.45\textwidth]{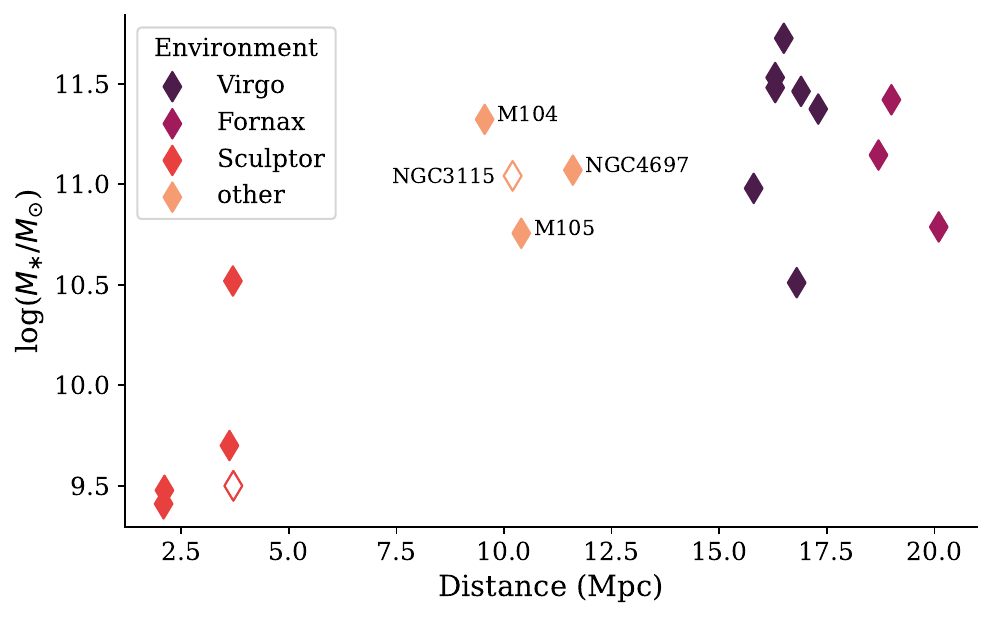}
    \caption{Distances and stellar masses of the GOLD sample. Different colours show different environments. NGC\,247 and NGC\,3115, which have not yet been observed, are shown by the empty symbols.}
    \label{fig:galaxy_sample}
\end{figure}

\section{Galaxy sample}
\label{sect:galaxy_sample}
At the time of writing, the galaxy sample covered in GOLD contains 19 target galaxies, with 17 galaxies fully observed\footnote{At the moment, observations for NGC\,247 and NGC\,3115 are not concluded, but scheduled in a filler programme.}. 
The galaxy sample is shown in Table \ref{tab:galaxy_sample} and Fig. \ref{fig:galaxy_sample}. The galaxies were chosen to span a range of masses ($M_\ast \sim 3 \times 10^{9} - 5\times 10^{11} M_\sun$), galaxy morphologies (late-type, early-type, and lenticular galaxies), and distances ($D \sim 2 - 20$ Mpc), and in the future we aim to add additional targets.

GOLD covers five star-forming galaxies in the Sculptor group (NGC\,253, NGC\,300, NGC\,7793, NGC\,247, and NGC\,55), which are also the closest galaxies in the sample, with distances of $D = 2 - 4$ Mpc. For these galaxies, high-quality distance measurements based on Cepheids and TRGB are available \citep{Anand2021}, but relatively little is known about the GC systems of Sculptor group galaxies due to their large extent on the sky and low systemic velocities, both of which make identification of GCs difficult (but see e.g. \citealt{Beasley2000, Olsen2004, Nantais2006, Cantiello2018, Fielder2025} for photometric and spectroscopic studies of GCs in the Sculptor group).

Twelve galaxies in our sample were selected to overlap with the JWST TRGB-SBF project \citep{Anand2024_Fornax, Anand2025_Virgo, Jensen2025}, an ongoing effort to derive TRGB and SBF distances to nearby galaxies using JWST NIRCam imaging to establish an alternative distance ladder for measurements of the Hubble constant, $H_0$. These are massive early-type galaxies mainly located in the Fornax and Virgo galaxy clusters, with distances ranging from $D \sim 10 - 20$ Mpc. Most of these galaxies have well-studied GC systems with both ground- and space-based imaging (e.g. \citealt{Jordan2009, Jordan2015, Cantiello2020, Powalka2016}), as well as low-resolution spectroscopy (e.g. SLUGGS data is available for M60, M84, M87, and NGC\,4697). Due to the large physical size of the FLAMES field of view (25\arcmin\,diameter), the pointings on some of these galaxies also include other galaxies by chance (e.g. NGC\,4647, which is close to M60, or NGC\,3384, close to M105).

Additionally, we have included the two lenticular galaxies M\,104 (NGC\,4594, the Sombrero galaxy) and NGC\,3115, which are at a similar distance of $\sim 10$ Mpc. We presented the FLAMES data of M\,104 in \cite{Fahrion2025b} and provided a distance measurement based on the GCVD approach in \cite{Fahrion2026a} as a pilot for the GOLD project. NGC\,3115 has a well-studied GC system (e.g. \citealt{Forbes2017, Dolfi2020, Dou2025}). It was also included in the sample but has not yet been observed.

\section{Data}
\label{sect:data}
Our spectroscopic observations were carried out with the GIRAFFE spectrograph of the FLAMES instrument at ESO's VLT between September 2024 and May 2026. We aim to achieve a minimum S/N $=$ 10 for sources with $V = 21$ mag to obtain precision velocity dispersion measurements with uncertainties of $< 2$ km s$^{-1}$, which could be reached in five hours per field. 

\subsection{Cluster selection strategy}
We aim to observe as many bona fide GCs as possible per galaxy using available photometric and spectroscopic GC catalogues as inputs. However, the depth, spatial extent, and quality of these literature catalogues vary strongly between the target galaxies. While the GC systems of some of our target galaxies have been studied extensively with both photometry and spectroscopy (e.g. M87 or NGC\,1399), for other galaxies only limited information is available. This is especially true for the galaxies in the Sculptor group (NGC\,253, NGC\,300, NGC\,247, NGC\,55, and NGC\,7793), for which only very limited photometric and almost no spectroscopic GC catalogues exist (but see \citealt{Cantiello2018} or \citealt{Olsen2004}). We describe the available photometric and spectroscopic GC catalogues that we used to prepare the observations in Appendix \ref{app:literature_catalogues}. From these, we selected GC candidates for observations following an iterative process for assigning priorities for fibre assignments.

The highest priority was assigned to spectroscopically confirmed GCs as well as \textit{Hubble} Space Telescope (HST)-based photometric candidates with $V < 21$ mag. Given that HST can marginally resolve GCs in our target galaxies, these are secure members. The second priority was given to spectroscopically confirmed or HST-selected GCs with $21 < V < 22$, together with ground-based photometric candidates brighter than $V < 21$ mag. Fainter GC candidates were given lower priority. Lastly, the input catalogues were extended to include candidates from Gaia DR3 \citep{GaiaCollaboration2023}, using magnitudes, colour, \textsc{astrometric\_excess\_noise}, and \textsc{phot\_bp\_rp\_excess\_factor} as selection parameters \citep{Voggel2020}.
This priority list was then used as input to the \textsc{fposs} tool\footnote{\url{https://www.eso.org/sci/observing/phase2/SMGuidelines/FPOSS.html}} to prepare the FLAMES observations, filling as many fibres as possible. Finally, at least 12 fibres were placed on empty sky regions to ensure a good sky subtraction.

We note that, by observing the brighter clusters, our sample may not be fully representative of GCs in the MW, depending on the distance of the galaxy. Consequently, our sample also contains very massive GCs ($> 10^6 M_\sun$) as well as UCDs. While UCDs are another type of compact stellar system that shares many properties of massive GCs, they may also consist of remnant stellar nuclei of disrupted satellite galaxies (e.g. \citealt{Hilker1999, Drinkwater2000, Wang2023}). When building our catalogues to prepare observations, we therefore also included known UCDs and UCD candidates from \cite{Liu2020} and \cite{Fahrion2019}.

\begin{table}[]
    \centering
    \caption{Observation details.}
    \begin{tabular}{c c c c c}\hline\hline
    Galaxy & $N_\text{fields}$ & $t_\text{exposure}$ & $<$seeing$>$ & Programme ID \\ 
    & & (s) & (arcsec) & \\ 
    (1) & (2) & (3) & (4) & (5) \\ \hline
M104 & 2 & 30030 & 0.75 & 114.274W.001 \\
M49 & 3 & 40950 & 0.72 & 116.28QP.001 \\
M60 & 1 & 13650 & 0.93 & 116.28QP.001 \\
M84 & 1 & 14541 & 0.72 & 116.28QP.001 \\
M86 & 1 & 13650 & 0.53 & 116.28QP.001 \\
M87 & 3 & 45045 & 0.79 & 116.28QP.001 \\
M89 & 1 & 13650 & 0.71 & 116.28QP.001 \\
M105 & 1 & 19110 & 0.92 & 116.28QP.001 \\
NGC\,4636 & 1 & 13650 & 0.52 & 116.28QP.001 \\
NGC\,4697 & 1 & 16380 & 0.68 & 116.28QP.001 \\
NGC\,1399 & 1 & 13650 & 0.47 & 116.28QP.001 \\
NGC\,1404 & 1 & 13650 & 0.61 & 116.28QP.001 \\
NGC\,1380 & 1 & 16380 & 0.93 & 116.28QP.001 \\
NGC\,253 & 2 & 28665 & 0.78 & 114.274W.001 \\
NGC\,300 & 4 & 51870 & 0.64 & 115.27YL.001 \\
NGC\,55 & 2 & 24570 & 0.49 & 115.27YL.001 \\
NGC\,7793 & 3 & 52544 & 0.82 & 115.27YL.001 \\
\hline
    \end{tabular}
    \label{tab:obs_details}
    \tablefoot{(1) Galaxy name, (2) number of FLAMES fields, (3) total exposure time across all fields, (4) average seeing, (5) Programme ID.}
\end{table}

\subsection{FLAMES observations}
All data were obtained with the GIRAFFE spectrograph of the FLAMES instrument, using the MEDUSA fibre positioner that allowed us to allocate up to 135 fibres per field in a field of view with 25\arcmin\, diameter (see App. \ref{app:pointing_setup} for an example observational setup). Each MEDUSA fibre spans an aperture of 1.2\arcsec, corresponding to physical sizes of between 11.6 pc (at $D=2$ Mpc) and 116 pc (at $D=20$ Mpc). The fibres therefore cover most of the GCs' light, even for the closest galaxies, but we shall derive aperture corrections in future works to correct any velocity dispersions (see \citealt{Beasley2025, Fahrion2026a}). We designed the observing strategy to cover every field by ten individual exposures of 1365 seconds each, distributed over five one-hour observing blocks per field. In some cases, observing blocks went out of the seeing constraints during observations or had to be repeated due to other reasons, which leads to slightly varying final exposure times per field (see Table \ref{tab:obs_details}). We also included exposures taken under worse than requested seeing constraints, as they led to improving signal-to-noise in the final spectra. Few corrupted exposures, such as for example those where guiding failed, were discarded.

We used the HR21 (875nm) grating, which covers the wavelength range from 8483 to 8900 \AA\,at a resolution of $R=18,000$ sampled at 0.05 \AA\,pix$^{-1}$. This grating includes the CaT comprising three Ca \textsc{II} lines at 8498.02, 8542.09, and 8662.14 \AA. This choice was motivated by the high throughput of the HR21 grating, and the fact that the CaT lines are prominent spectral features in GCs, mostly stemming from red giant branch and asymptotic branch stars. The contribution of giant stars to the continuum flux at 8500\AA\ in integrated spectra is approximately $\sim80$\% for a 12 Gyr, [Fe/H] = $-1.0$ dex stellar population typical of GCs \citep{Vazdekis2010}. The contribution to the CaT lines themselves is even greater, reaching 90\%, since the CaT is significantly stronger in giants than dwarf stars of a comparable temperature.
The CaT is also an excellent metallicity indicator, as the CaT line strength increases with metallicity for metallicities [M/H] $< -0.5$ dex \citep{ArmandroffZinn1988, Diaz1989, Foster2010, Foster2011, Usher2012, Sakari2022}. However, as both empirical comparisons and stellar population models show, the CaT feature saturates at higher metallicities due to the growth of titanium- and vanadium oxide (TiO, VO) molecular features \citep{Vazdekis2003, Usher2012}. While a majority of GCs is expected to lie below this threshold of $[M/H] < -0.5$ dex as measurements of extragalactic GC systems show (e.g. \citealt{Usher2019, Villaume2019, Fahrion2020b}), the availability of numerous iron lines may also enable more direct metallicity measurements. Additionally, because the CaT is sensitive to surface gravity, it may also provide constraints on the stellar initial mass function (IMF), offering an independent tracer of past star formation activity \citep{Vazdekis2003, Cenarro2004}. We investigate this aspect further using our data (see also Sect. \ref{sect:Chemistry}).

\begin{figure*}
    \centering
    \includegraphics[width=0.90 \textwidth]{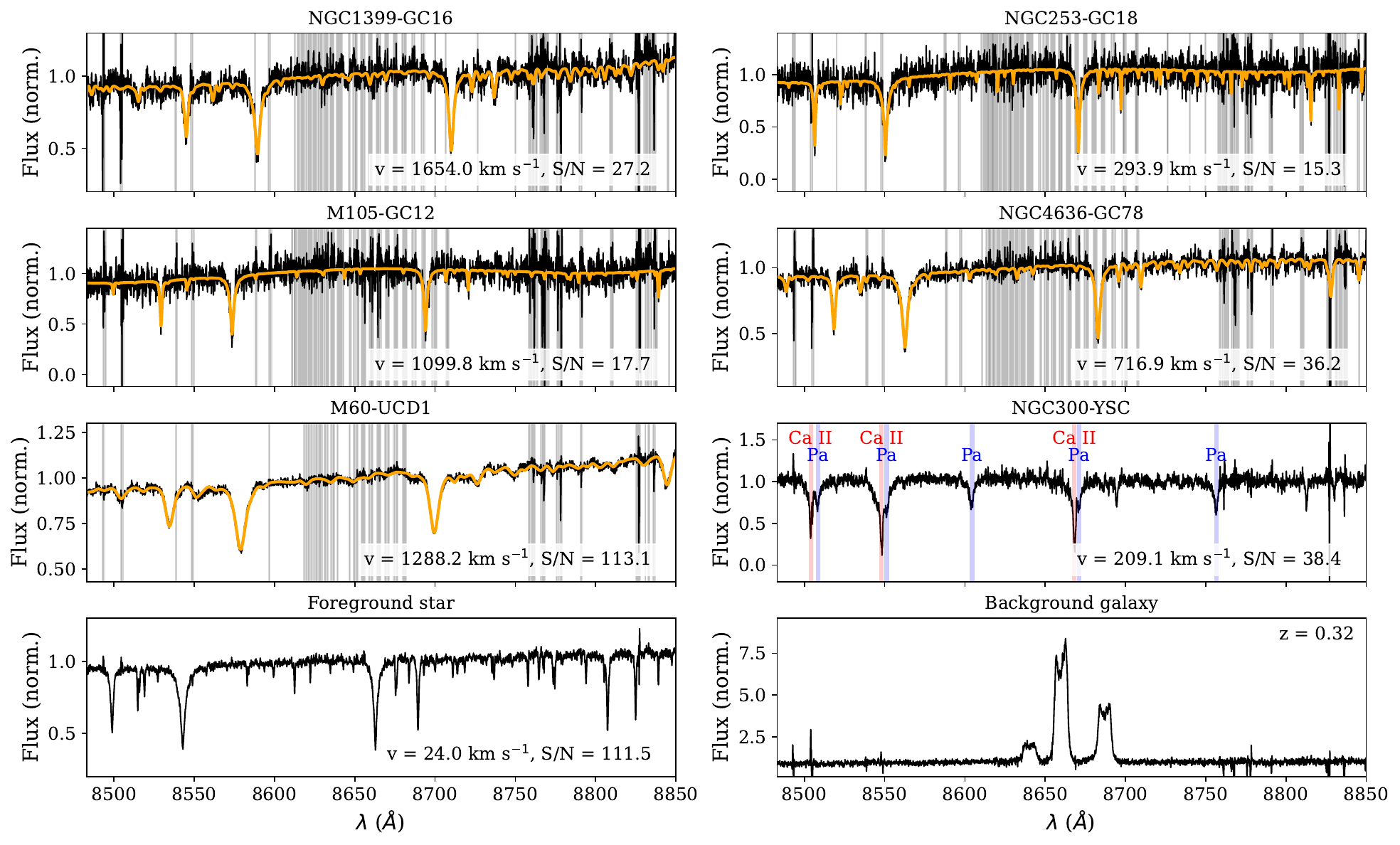}
    \caption{Example spectra FLAMES/GIRAFFE spectra. Top four panels show GC spectra (in black) from different galaxies and at different S/N, each fitted with \textsc{pPXF} (shown in orange). Regions with sky line residuals masked from the fit are shown as shaded grey areas. Third panel on the left shows the spectrum of M60-UCD1 as an example of an UCD spectrum in our sample. Other panels show the spectrum of a younger star cluster candidate in NGC\,300 with clearly visible Paschen absorption, an example foreground star spectrum, and the spectrum of a background galaxy with double-peaked H$\alpha$ and [NII] lines.}
    \label{fig:spectra}
\end{figure*}

Beyond the CaT lines themselves, the spectral region from $\sim 8500$ to 9000 \AA\,offers additional absorption lines that can be used in stellar population analyses. For old stellar populations, several Fe I lines are present in this spectral window, in addition to absorption lines from Ti I, Si I, Mg I, and Al I (\citealt{Carretta2012, Contursi2021}; see Sect. \ref{sect:Chemistry}). In younger stellar populations, strong absorption from lines of the hydrogen Paschen series are visible (e.g. see the spectrum of a younger cluster in Fig. \ref{fig:spectra} or \citealt{Usher2019_WAGGS}).

\subsection{Data reduction pipeline}
We reduced the FLAMES GIRAFFE data following the same steps as described in \cite{Fahrion2025b} and \cite{Fahrion2026a}, using a custom wrapper around the \textsc{esorex} command line utility for running the GIRAFFE reduction recipes\footnote{\url{https://www.eso.org/sci/software/pipelines/giraffe/giraffe-pipe-recipes.html}}. The data reduction steps included bias and flat fielding corrections, and a wavelength calibration. Following the FLAMES instrument manual, we repeated the wavelength calibration to obtain an improved wavelength solution. To extract the spectra, we selected the \textsc{OPTIMAL} method \citep{Horne1986}. 

In a second step, we subtracted the sky for each exposure independently using a custom approach as described in detail in \cite{Fahrion2026a}. In summary, we created a library of sky spectra using the sky fibres and then used the penalised pixel fitting method (\textsc{pPXF}; \citealt{Cappellari2004, Cappellari2017, Cappellari2023}) to fit each spectrum with both the sky spectra and red giant branch star models from the PHOENIX library \citep{Husser2013}. This approach allows us to scale the sky spectra appropriately instead of just subtracting a master sky spectrum. For each exposure, the spectrum with and without sky subtraction was saved.

For the final combination of the spectra of each field, we accounted for the barycentric velocity of each exposure and applied a signal-to-noise-weighted median combination. We also saved a version of the spectra without any sky subtraction to subsequently identify regions with strong sky residuals that can be masked when fitting the spectra (see Fig. \ref{fig:spectra}).

\section{Spectroscopic analysis}
\label{sect:methods}
We describe here how the FLAMES GIRAFFE spectra were analysed. Figure \ref{fig:spectra} shows several example spectra of GCs and other sources.

\subsection{Radial velocities}
In total, we obtained 2701 spectra for the 17 target galaxies. To categorise the spectra, we inspected them individually. About 933 have either a flat, featureless spectrum or are too noisy to identify the CaT lines. We identified 42 background galaxies, mainly based on redshifted emission lines of H$\alpha$ together with [NII], or the [SII] doublet. The remaining sample of 1726 objects contains objects with CaT lines, either stemming from genuine GCs (or UCDs) or MW foreground stars. From this sample, we removed 25 objects due to either very broad or asymmetric CaT lines, which likely were caused by light from the host galaxy contaminating the fibres significantly.

The remaining 1701 objects were fitted with \textsc{pPXF}, which uses model spectra to obtain the parameters of the line-of-sight velocity distribution (LOSVD) such as the LOS mean velocity and velocity dispersion (see e.g. \citealt{Fahrion2026a}). The \textsc{pPXF} code can also be used to obtain mean ages and metallicities when stellar population models are used to fit the spectra, but in this paper we only focus on the LOS velocities to get a first idea of the GC sample contained in GOLD.

We used the high-resolution stellar models from \cite{Coelho2014} to fit the FLAMES spectra. These provide stellar models at a resolution of $R = 20,000$, which we convolved to a FWHM = 0.54 \AA, the average instrumental resolution of our FLAMES data (see \citealt{Fahrion2026a}). In the fit, we used additive polynomials of degree 4 and multiplicative polynomials of degree 1 to minimise template mismatch, based on our testing in \cite{Beasley2025}. We used the wavelength range from 8485 to 8850 \AA. Since there are still sky residual lines present in the spectra, we created custom masks by excluding every wavelength bin where the normalised sky spectrum exceeds 1.2. We settled on this choice after some testing described in \cite{Beasley2025}. To obtain an estimate of the uncertainties on the LOS velocities, we performed a bootstrapping approach. Following the first fit, the best-fit model was perturbed by randomly drawing from the residual in each wavelength bin, creating 100 representations of the spectrum. These were then fitted to obtain the LOS velocity and its uncertainty as the median and standard deviation of the resulting distribution, reported in Table \ref{tab:losv_catalogue}.

In the future, when measuring velocity dispersions, we shall use the large GOLD dataset to test the steps described here, including the choice of the model spectra and the instrumental resolution, in more detail. However, we find that the LOS velocities are robust against changes to the \textsc{pPXF} fitting setup. 

\subsection{Velocity distributions}
\begin{figure*}
    \centering
    \includegraphics[width=0.85\textwidth]{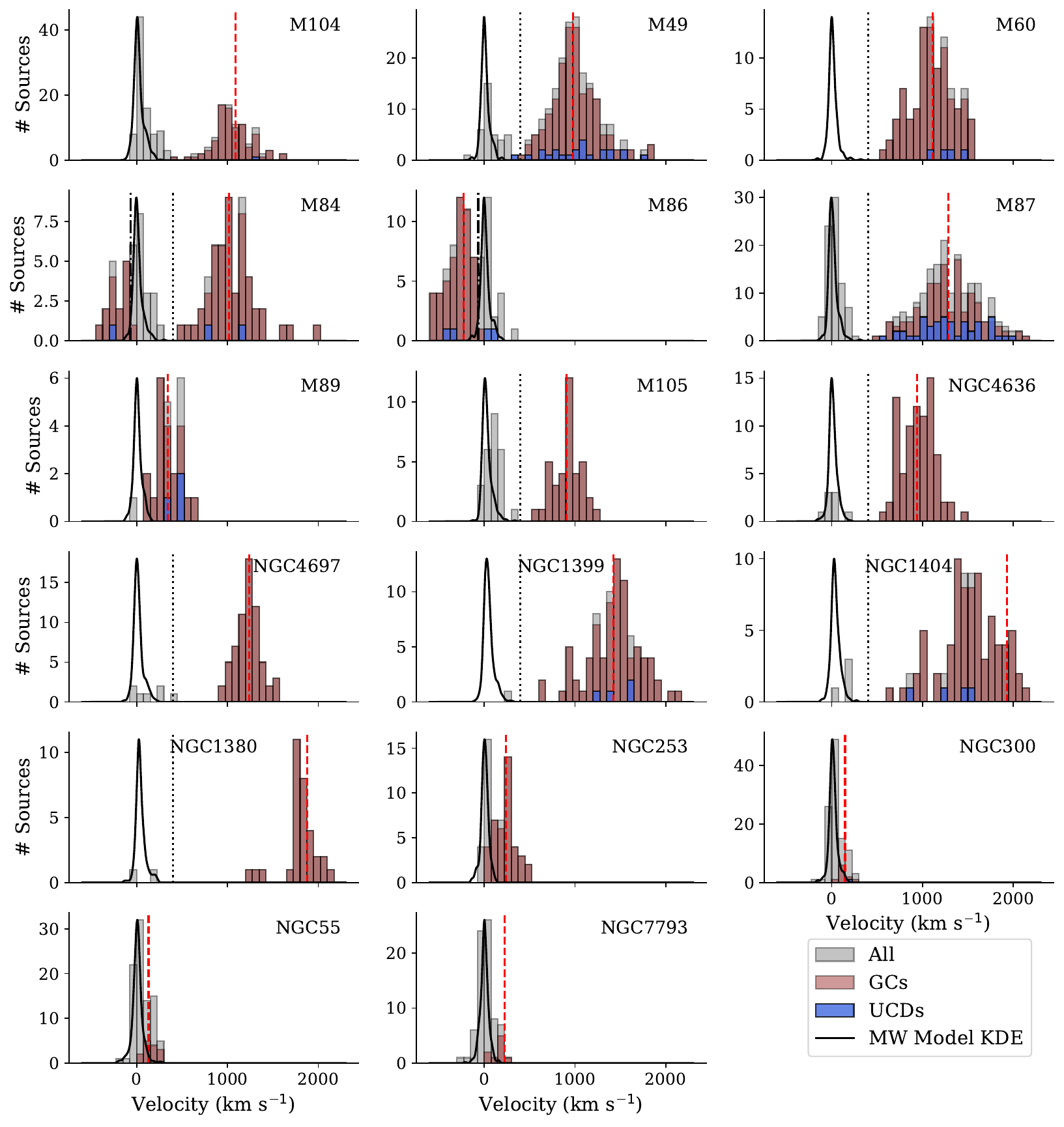}
    \caption{Histograms of LOS velocities of GOLD objects with visible CaT lines in their spectrum (grey), and selected GCs (red) and UCDs (blue). The black lines show the velocity distribution of MW foreground stars, obtained from the Besan\c{c}on model for each galaxy position, scaled to the peak of the histogram for visualisation. The dashed red line indicates the systemic velocity of each galaxy. The dash-dotted and dotted black lines indicate velocity cut-offs at $-65$ and 400 km s$^{-1}$, respectively, which were chosen to select the GC sample as described in the text.}
    \label{fig:velocity_distributions}
\end{figure*}

We present the velocity distributions of all objects with CaT lines in Fig. \ref{fig:velocity_distributions}, separated by target galaxy. In most cases, there is a bimodal distribution visible, illustrating the genuine GC population of the target galaxy and contamination from MW foreground stars. To show the expected velocity distribution of MW stars, we included model velocities from the Besan\c{c}on MW model \citep{Robin2003, Czekaj2014, Lagarde2017}, obtained for each galaxy's position. The MW model distribution and GC velocities are also compared to the systemic velocity of the host system. While in many galaxies the GCs can be easily separated from MW stars based on their velocity alone, the separation can be challenging in some of the target galaxies with low systemic velocities (e.g. the Sculptor galaxies or M86).

\begin{figure*}
    \centering
    \includegraphics[width=0.85\textwidth]{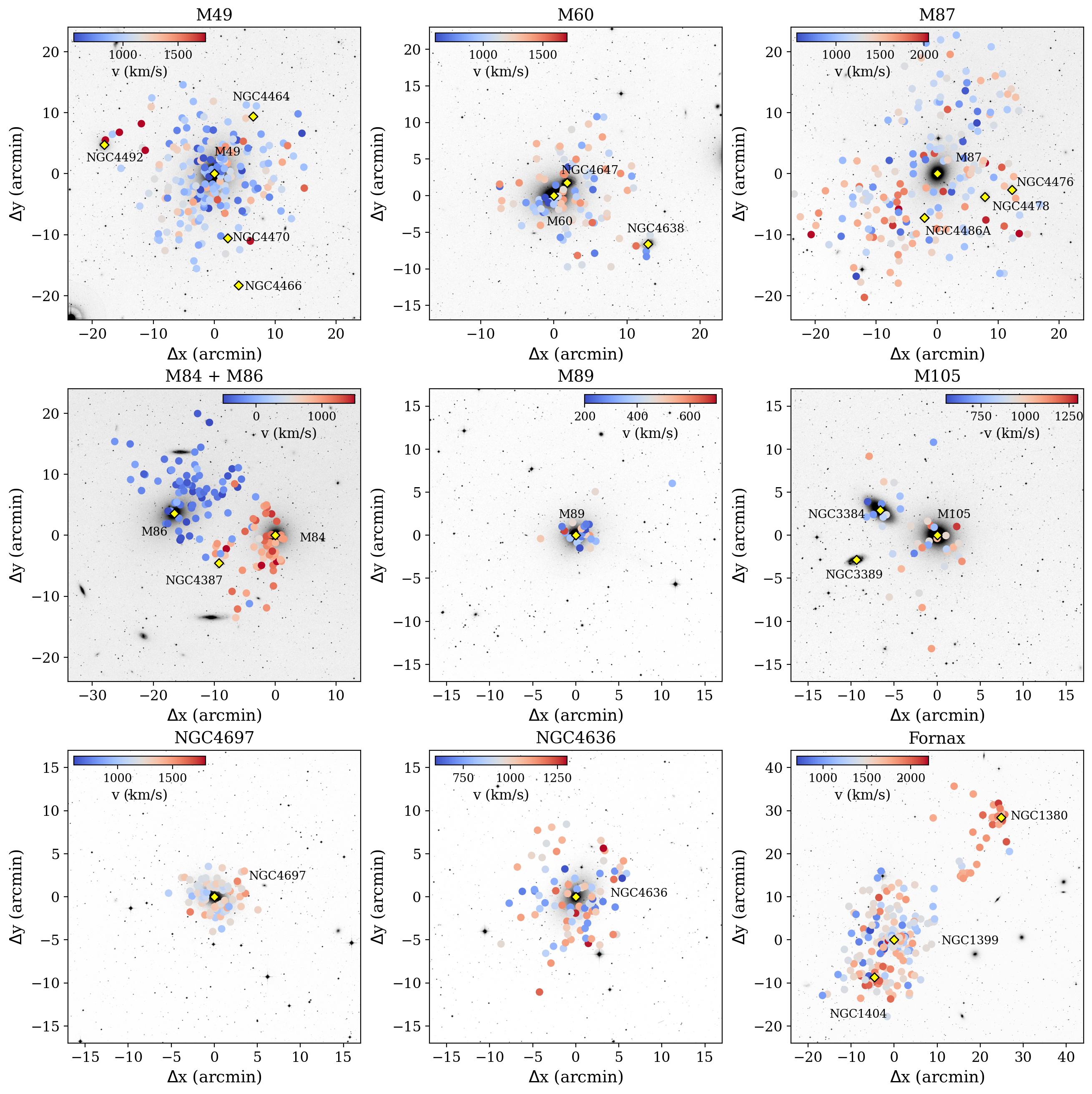}
    \caption{Spatial distribution of velocities. Each panel shows a DSS2 image in grey scale of a different target galaxy or environment. Circles indicate our GCs and UCDs, colour-coded by their LOS velocities. In each panel, the range of the colour bar was chosen to demonstrate the range of LOS velocities around the systemic velocity of the host galaxies. Yellow diamonds indicate prominent galaxies in the field of view.}
    \label{fig:spatial_velocities}
\end{figure*}

In several cases, Fig. \ref{fig:velocity_distributions} also shows multiple velocity peaks due to other nearby galaxies that were covered in the same FLAMES field. To illustrate this, we show the spatial distributions of LOS velocities in Fig. \ref{fig:spatial_velocities} on top of galaxy images from the Digitized Sky Survey 2 (DSS2)\footnote{\url{https://archive.eso.org/dss/dss}}. While M89, NGC\,4697, and NGC\,4636 appear to be rather isolated, the fields around M49, M87, and in the Fornax cluster contain several other galaxies. Additionally, M84 and M86 are only separated by 16\arcmin\, on the sky, which explains why the velocity distribution for M84 shows some low-velocity sources that likely are GCs belonging to M86. Moreover, M60 ($v_\text{sys} = 1110$ km s$^{-1}$ according to NED) lies very close to the spiral galaxy NGC\,4647 ($v_\text{sys} = 1409$ km s$^{-1}$), and M105 ($v_\text{sys} = 907$ km s$^{-1}$) lies close to NGC\,3384 ($v_\text{sys} = 704$ km s$^{-1}$), and our samples likely contain GCs from both galaxies in these cases. 

\begin{table}[]
    \centering
    
    \caption{Numbers of selected GCs and UCDs per field.}
    \begin{tabular}{c c c} \hline\hline
      Galaxy   &  $N_\text{GCs}$ & $N_\text{UCDs}$\\ \hline
       M104 & 86 & 1 \\
    M49 & 166 & 23 \\
    M60 & 90 & 4 \\
    M84 & 66 & 4 \\
    M86 & 54 & 4 \\
    M87 & 133 & 49 \\
    M89 & 21 & 3 \\
    M105 & 38 & 0 \\
    NGC\,4636 & 81 & 0 \\
    NGC\,4697 & 65 & 0 \\
    NGC\,1399 & 80 & 6 \\
    NGC\,1404 & 77 & 6 \\
    NGC\,1380 & 32 & 0 \\
    NGC\,253 & 40 & 0 \\
    NGC\,300 & 10 & 1 \\
    NGC\,55 & 14 & 0 \\
    NGC\,7793 & 9 & 0 \\ \hline
    \end{tabular}
    \label{tab:GC_UCD_numbers}
\end{table}
\subsection{Selecting the initial GC sample}
\label{sect:selecting}
Given the complex velocity distributions of our sample, we used a multi-step process to select our initial GC sample. However, we note that this sample may be subject to change in the future as especially our planned distance measurements might require a more restrictive selection of GCs, and potentially measured velocity dispersions could be used to identify GCs. In this work, we simply aim to distinguish between GCs and foreground stars, regardless of the host galaxy.

We started with simple selections for the cases in which the GCs can be easily separated in velocity from foreground stars based on their velocity. For the ten FLAMES fields of M49, M60, M84, M87, M105, NGC\,4636, NGC\,4697, NGC\,1399, NGC\,1404, and NGC\,1380, we simply categorised every object with a velocity greater than 400 km s$^{-1}$ as a GC. This cut was chosen to remove even high-velocity MW stars, but appears to be well justified from the gap in the velocity distributions (see Fig. \ref{fig:velocity_distributions}). Sources that could be matched to the UCD catalogues by \cite{Liu2020} and \cite{Fahrion2019} were instead classified as UCDs (see App. \ref{app:literature_catalogues}). This simple selection yields 813 GCs and 92 UCDs in these ten galaxies. 

For M104, we used the selection presented in \cite{Fahrion2025b}, which adds another 86 GCs and one UCD \citep{Hau2009}. 
For M89, we started with the same velocity selection, $v > 400$ km s$^{-1}$, and then added sources that could be matched to the HST ACS catalogues of \cite{Jordan2009}. As HST is able to marginally resolve typical GCs at a distance of the Virgo and Fornax galaxy clusters, such HST catalogues are expected to have only modest contamination. We therefore classified all objects as GCs that fulfil $pGC > 0.5$, where $pGC$ gives a measure of the probability that a source is a GC \citep{Jordan2009, Jordan2015}, and classified sources as UCDs that matched with the UCD catalogue by \cite{Liu2020}. In total, 21 GCs and 3 UCDs were added here.

For M86, we followed a similar procedure as for M89, adding GCs and UCDs from the catalogues of \cite{Jordan2009} and \cite{Liu2020}, respectively. Then, we categorised every object with a $v < -65$ km s$^{-1}$ as a GC, as only 5\% of MW stars are expected to have such a blueshifted velocity following the Besan\c{c}on model at this position. The same cut was applied to the GCs in the M84 field due to the close proximity of both galaxies in the sky. In this way, we added 69 GCs and 4 UCDs.  

The selection of GCs is the most challenging for the Sculptor galaxies due to their low but positive systemic velocities, and low stellar masses, which reduces the number of expected GCs (e.g. \citealt{BurkertForbes2020}). However, the Sculptor galaxies are also the closest in our sample, and any star cluster can be visually identified as such from high-resolution space-based imaging if available. In the first step to collect GCs, we went to the MAST archive and manually inspected all available HST images that contain at least one source with a velocity of $> 50$ km s$^{-1}$, which is larger than 90\% of expected foreground stars. In this way, we were able to identify 15 GCs in NGC\,253 (see also \citealt{Cantiello2018}, who used HST images to check against their ground-based GC catalogue), one cluster in NGC\,7793, one in NGC\,300, and four in NGC\,55.

In the next step, we made use of Gaia DR3 photometry \citep{GaiaCollaboration2023}, where available, removing those GC candidates that show \textsc{phot\_bp\_rp\_excess\_factor} $< 1.3$ or \textsc{astrometric\_excess\_noise} $<0.5$. For spatially extended sources such as GCs at a distance of the Sculptor group galaxies, these parameters are expected to be elevated compared to unresolved sources (see \citealt{Voggel2020}). Other candidates with velocities of $>$ 50 km s$^{-1}$ but no Gaia photometry were kept. This selection procedure yielded 40 GCs in NGC\,253, 9 GCs in NGC\,7793, 14 GCs in NGC\,55, and 10 GCs and one UCD \citep{Fielder2025}\footnote{\cite{Fielder2025} does not consider NGC300-GCF25 as a UCD, but rather a GC. We placed it in our UCD sample due to its large size $R_\text{eff} > 10$ pc, which is substantially above the typical size of GCs ($\sim 3$ pc).} in NGC\,300.

In total, our initial sample consists of 1062 GCs and 101 UCDs (see Table \ref{tab:GC_UCD_numbers}). We make the catalogue of velocities for these sources publicly available, as described in App. \ref{app:vel_catalogue}.

\section{Properties of the GC sample}
\label{sect:properties}
In this section, we aim to present our initial GC sample and its properties across the different galaxies. 
\subsection{Signal-to-noise and velocity uncertainties}
\begin{figure}
    \centering
    \includegraphics[width=0.43\textwidth]{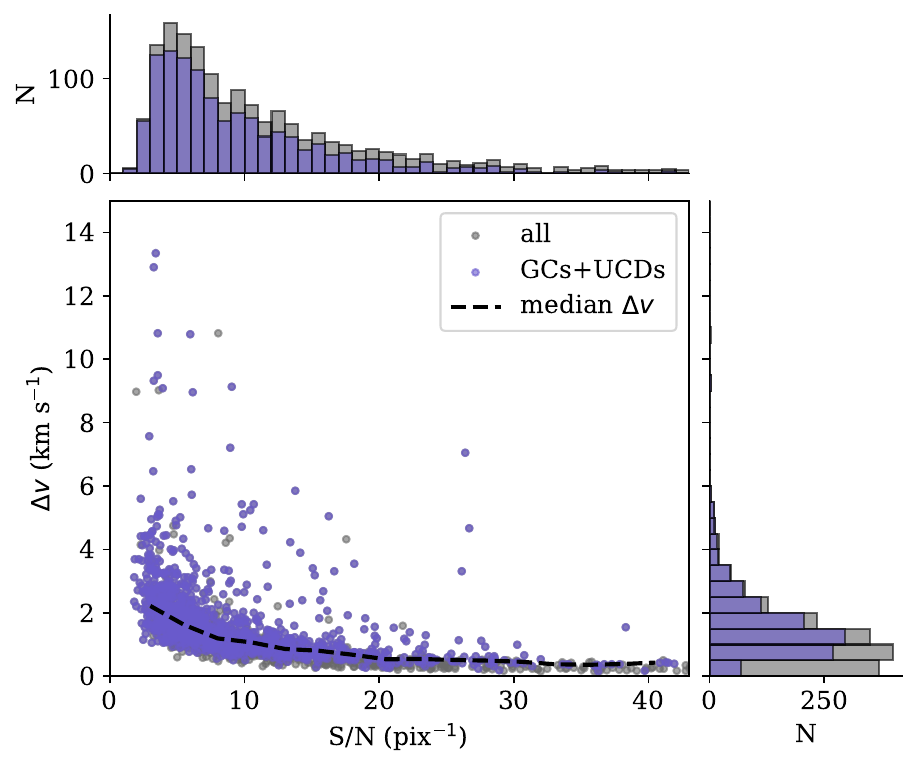}
    \caption{Signal-to-noise per pixel versus LOS velocity uncertainty. Grey symbols refer to all 1701 sources with measured velocities including foreground stars; purple symbols show GCs and UCDs from our selection. The dashed line shows the median velocity uncertainty. We note that there are a 21 sources (13 GCs) with $\Delta v$ > 15 km s$^{-1}$ and S/N $<$ 3 pix$^{-1}$ that are outside of the plotting range. Also, there are 159 sources with S/N $>$ 45 pix$^{-1}$; 13 of them are GCs and UCDs.}
    \label{fig:SNR_vs_dv}
\end{figure}

We show the distribution of S/N (per pixel)\footnote{At a sampling of 0.05 \AA\,pix$^{-1}$, the S/N per pixel can be converted to S/N per \AA\,using a factor of $\sqrt{20}$.} and the derived LOS velocity uncertainties in Fig. \ref{fig:SNR_vs_dv} for all sources and the selected GCs and UCDs.
At S/N $\sim 5$ pix$^{-1}$ ($V \sim 21.3$ mag), which roughly corresponds to the peak of the S/N distribution shown in Fig. \ref{fig:SNR_vs_dv}, the median velocity uncertainty is $\Delta v \sim 2$ km s$^{-1}$, roughly an order of magnitude smaller than for typical lower-resolution GC spectroscopy. At S/N = 10 pix$^{-1}$ ($V \sim 20.4$ mag), the median velocity uncertainty reaches $\Delta v \sim 1$ km s$^{-1}$.

\subsection{Magnitudes and colours}
\label{sect:mags}
While all our GCs and candidates have high-resolution FLAMES spectra in common, the available photometry and spectroscopy from literature sources is heterogenous. As described in Sect. \ref{sect:data} and Appendix \ref{app:literature_catalogues}, we used various available catalogues to construct our sample, including both ground- and space-based photometric catalogues as well as other available spectroscopic samples. 
Of 1163 GCs and UCDs, 465 have available $V$-band magnitudes, 769 have $g$-band magnitudes available, and 556 have Gaia $G$-band magnitudes. In the future, the number of matches reported here will potentially increase further, especially when data from wide-field imaging missions such as Euclid, the Roman Space Telescope, or the Vera Rubin Telescope become available.

\begin{figure}
    \centering
    \includegraphics[width=0.45\textwidth]{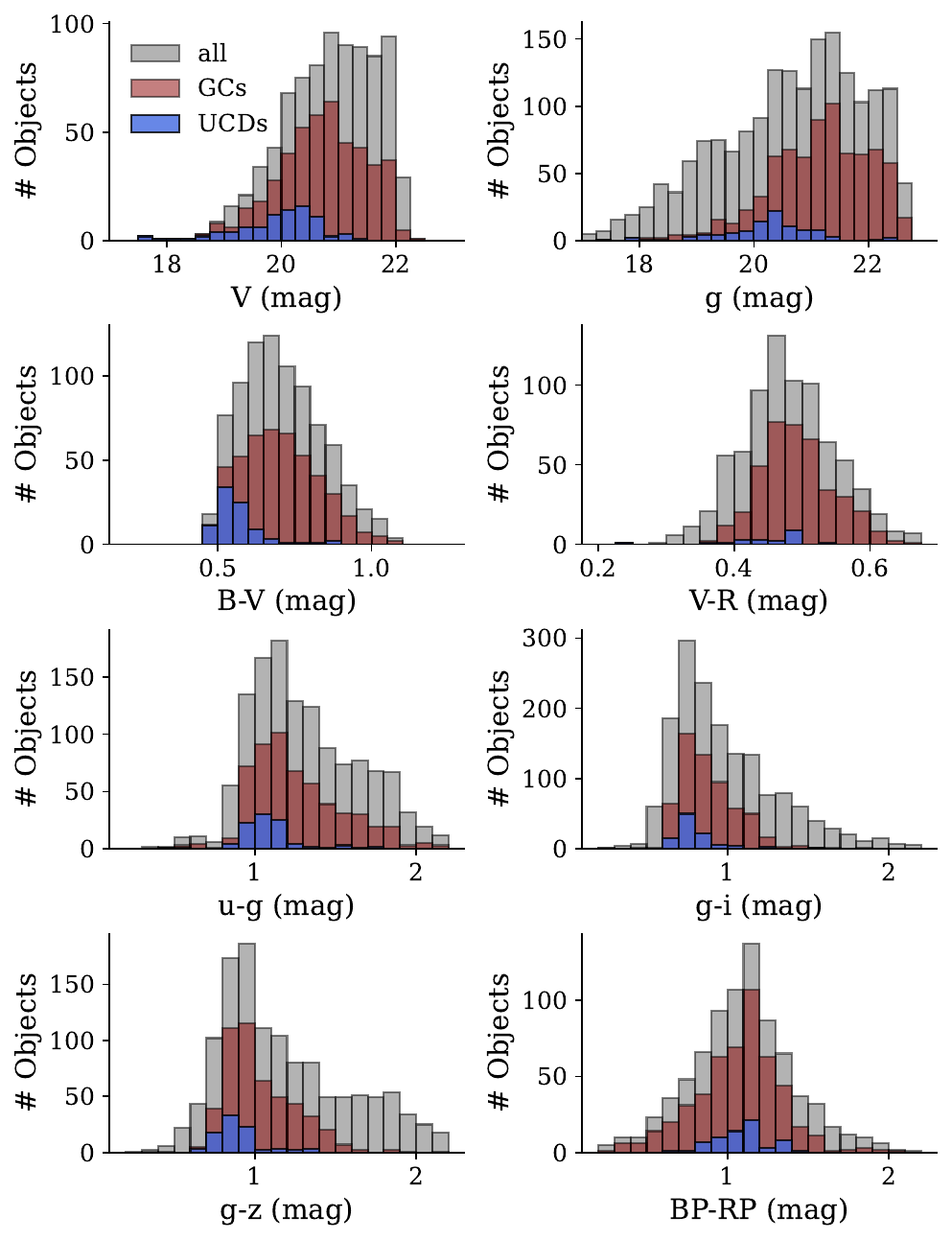}
    \caption{Histograms showing magnitude (top row) and colour distributions of our sample (bottom three rows). Grey histograms show all sources with measured velocities. Selected GCs are shown in red, UCDs in blue.}
    \label{fig:colour_dist}
\end{figure}

Figure \ref{fig:colour_dist} shows the distribution of $V$ and $g$-band magnitudes as well as colours in various filter combinations for all 1701 sources with measured velocities (grey) as well as our initial sample of GCs and UCDs selection (red and blue, respectively). By construction, given our selection criteria for fibre assignment, almost all sources are brighter than $V < 22$ mag, translating to $g \leq 22.4$ mag. The colour distributions of our initial GC+UCD sample are well defined and mostly unimodal, likely because our sample is dominated by very massive galaxies \citep{Peng2006, Harris2017b}. Additionally, the so-called `blue tilt' describes the finding that bright, blue GCs are typically redder than their fainter counterparts (e.g. \citealt{Harris2006, Spitler2006, Usher2018}), which further narrows the colour distributions of the bright GCs in our sample.

\subsection{Sizes}
\begin{figure}
    \centering
    \includegraphics[width=0.43\textwidth]{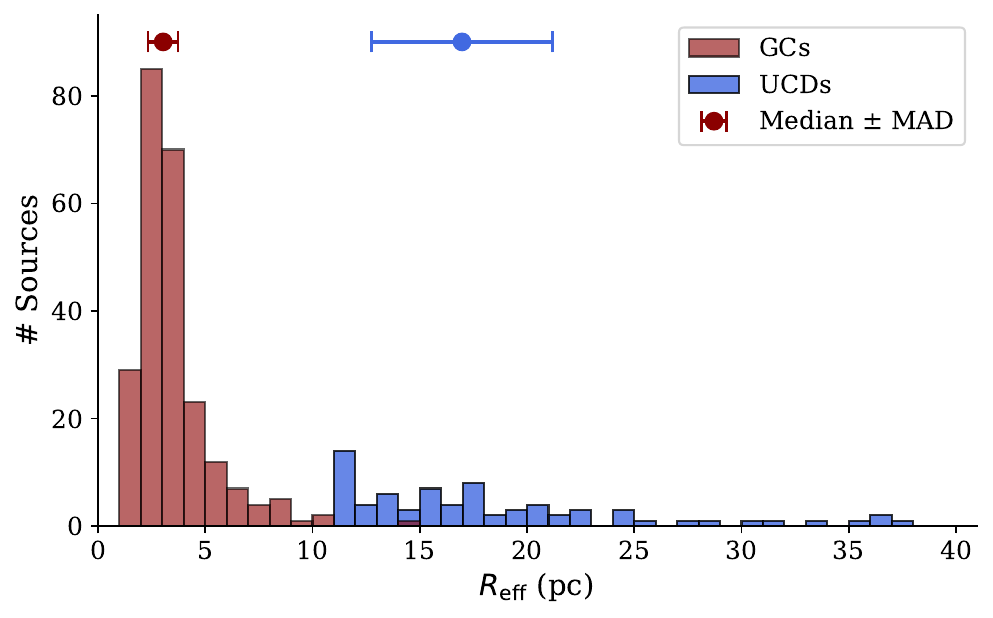}
    \caption{Effective radii of GCs (red) and UCDs (blue). The error bars refer to the median and median absolute deviation of each distribution.}
    \label{fig:sizes}
\end{figure}

Size measurements for compact objects such as GCs require high-resolution imaging data, which are only available for a subset of our sample. Cross-matching against available catalogues (see App. \ref{app:literature_catalogues}), we obtain measurements of effective radii of 239 GCs and 75 UCDs. The distribution of those are shown in Fig. \ref{fig:sizes}. 

The GCs have a narrow size distribution, with a median size of $R_\text{eff} = 3.04$ pc and a median absolute deviation of 0.69 pc. This is in agreement with findings from larger studies investigating the sizes of extragalactic GCs (e.g. \citealt{Masters2010}). These sizes also compare favourably to the MW size distribution, $R_\text{eff} = 3.19$ pc, and median absolute deviation = 1.14 pc (using the projected half-light radii from \citealt{Baumgardt2018}). In contrast, the UCDs in our sample are much more extended, reaching $R_\text{eff} \geq 10 - 30$ pc. One reason for this is that the selection of Virgo Cluster UCDs by \cite{Liu2020} utilised  ground-based magnitudes and size measurements to choose UCD candidates. However, UCDs in general tend to be larger than GCs.

\begin{figure*}
    \centering
    \includegraphics[width=0.93\textwidth]{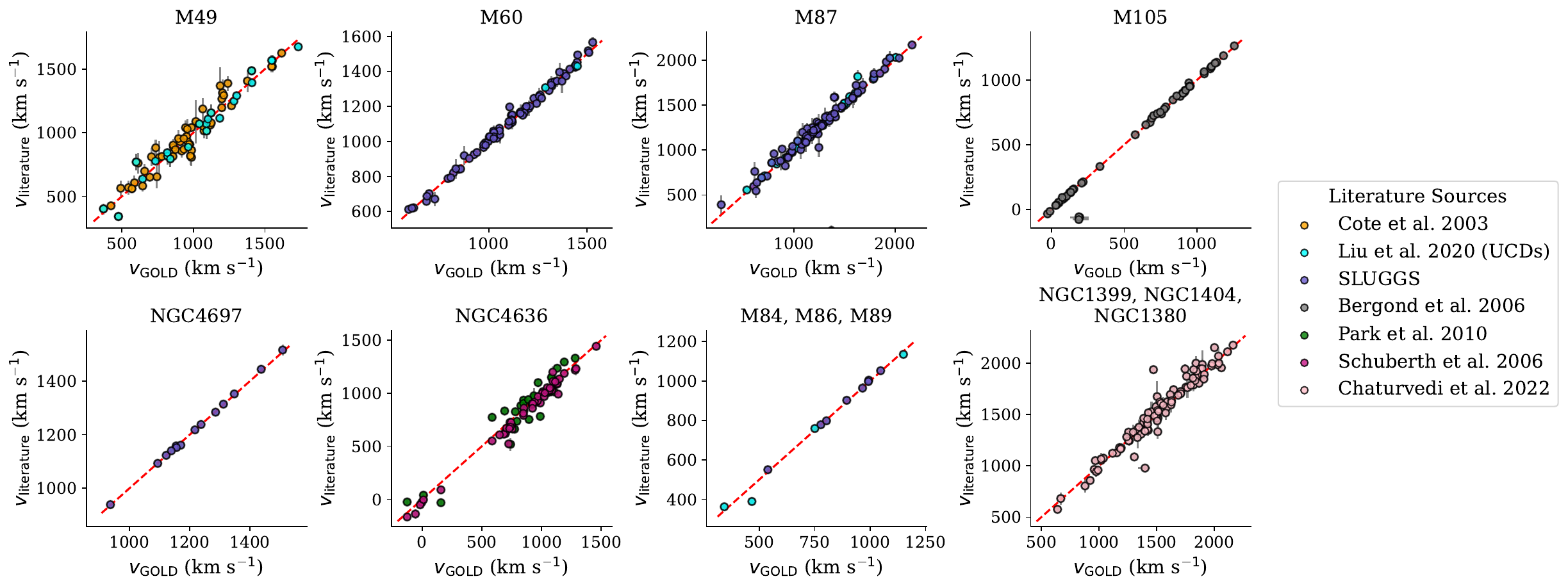}
    \caption{Comparison between LOS velocities from GOLD and literature sources in different galaxies (see App. \ref{app:literature_catalogues}). The coloured dots show velocities from \cite{Cote2003} in orange, and UCD velocities from \cite{Liu2020} in cyan, from the SLUGGS survey \citep{Forbes2017} in purple, from \cite{Bergond2006} in grey, from \cite{Park2010} in green, from \cite{Schuberth2006} in magenta, and from \cite{Chaturvedi2022} in pink. The dashed red line shows the one-to-one relation. Errors are shown where available; however, error bars are too small to be visible in many cases.}
    \label{fig:velocity_comparisons}
\end{figure*}
\subsection{Velocity comparisons}
With our target galaxies mostly being well-studied systems, other GC velocity catalogues exist that come from various sources (see App. \ref{app:literature_catalogues}). Of our sources, 630 have previously published velocities. Of those, 429 are in our GC sample, 70 are UCDs, and the remaining 131 sources are in our foreground star selection. Consequently, we present here the first velocity measurements for more than 600 GCs and 31 UCDs.

In Fig. \ref{fig:velocity_comparisons}, we plot our FLAMES-based velocities against these other catalogues, separated by galaxy or environment. As this figure shows, our velocities mostly agree very well with the literature catalogues, with the exception of a few outliers. Considering all matched sources that also have a velocity uncertainty available, we find that 91\% agree within $3\sigma$ of the combined uncertainties. For future work, we shall carefully inspect sources and their spectral fits that deviate from literature measurements, and shall exclude them in the subsequent analysis if necessary.
In some cases, for example for the M49 GC velocities from \cite{Cote2003}, there seems to be a broader scatter around the one-to-one relation. However, we note that our velocities have uncertainties of $\sim 1 - 2$ km s$^{-1}$, whereas the velocities from \cite{Cote2003} have uncertainties of between 20 and 60 km s$^{-1}$.

\section{Science with GOLD}
\label{sect:science}
In this section, we briefly discuss the science cases that we aim to address with the GOLD project.  

\subsection{Distances based on GCVD}
The primary goal of GOLD is to measure distances using the GCVD distance method, which relates the internal stellar velocity dispersion of GCs with their apparent magnitudes \citep{Beasley2024, Beasley2025, Fahrion2026a}. \cite{Beasley2024} demonstrated that the GCs of the MW and M31 show very similar $M_V - \sigma$ relations, and that by anchoring the MW GC distances via Gaia, distance uncertainties to other galaxies of $\sim3-5\%$ can be achieved. 

The GCVD relation is based on the Virial theorem that connects the luminous and dynamical mass of GCs and is therefore based on a well-understood theoretical framework. In principle, the GCVD approach provides distances competitive with other distance indicators such as the TRGB, Cepheids, and SBF. The approach even has several potential advantages over traditional methods:
\begin{itemize}
\item{Distances based on the GCVD method are independent of galaxy morphology. If the $M/L$ of GCs is approximately invariant across galaxy types, then the same distance measurement 
can be used across all types of galaxies that have GC populations. This is not the case for Cepheids (which require young stellar populations to be present in the galaxy), and varying underlying galaxy stellar populations may well systematically affect distances determined from the SBF and TRGB methods. With GOLD, we aim to test the universality of the GCVD across galaxy types and shall use available size measurements to directly constrain dynamical $M/L$ values in different host environments.}
\item{As ubiquitous objects, GCs are found in galaxies of all types, from low-mass dwarf galaxies to the most massive galaxies in the Universe, whereas SBF distances are best established in massive early-type galaxies and Cepheids require young populations. Additionally, GCs are predominantly found in the haloes of their host galaxies, and generally unaffected by crowding or intrinsic extinction. In contrast, extinction can affect especially distance measurements from Cepheids, and blending of sources can limit the accuracy of TRGB measurements.}
\item{The GCVD relation is primarily anchored to Gaia parallaxes (i.e. a geometric anchor). Cepheids, TRGB and SBF all require independent anchors, which may vary or depend on galaxy type.}
\end{itemize} 

In \cite{Fahrion2026a}, we applied the GCVD method to the FLAMES data of M104, as a first pilot study for GOLD. We found a distance of $9.0 \pm 0.3$ Mpc, comparable to other distance measurements (e.g. $9.1$ Mpc found by \citealt{Jensen2003} using SBF, or 9.6 Mpc derived from TRGB by \citealt{McQuinn2016}). As the other GOLD target galaxies were observed with the same setup, we expect to reach similar distance uncertainties, especially for the galaxies with similar or higher numbers of GCs.
In our pilot work of M104 \citep{Fahrion2026a}, we also started to explore how the scatter in the relation relates with properties of the GCs, such as the projected galactocentric distance, colours, or sizes. With the full GOLD sample and its rich auxiliary data, we aim to test in detail and potentially correct for any dependence of the GCVD relation with such properties. Our primary purpose is to explore whether the potential advantages of GCVD distances hold fast under scrutiny, and to derive independent distances to our target galaxies. 

\begin{figure*}
    \centering
    \includegraphics[width=0.85\textwidth]{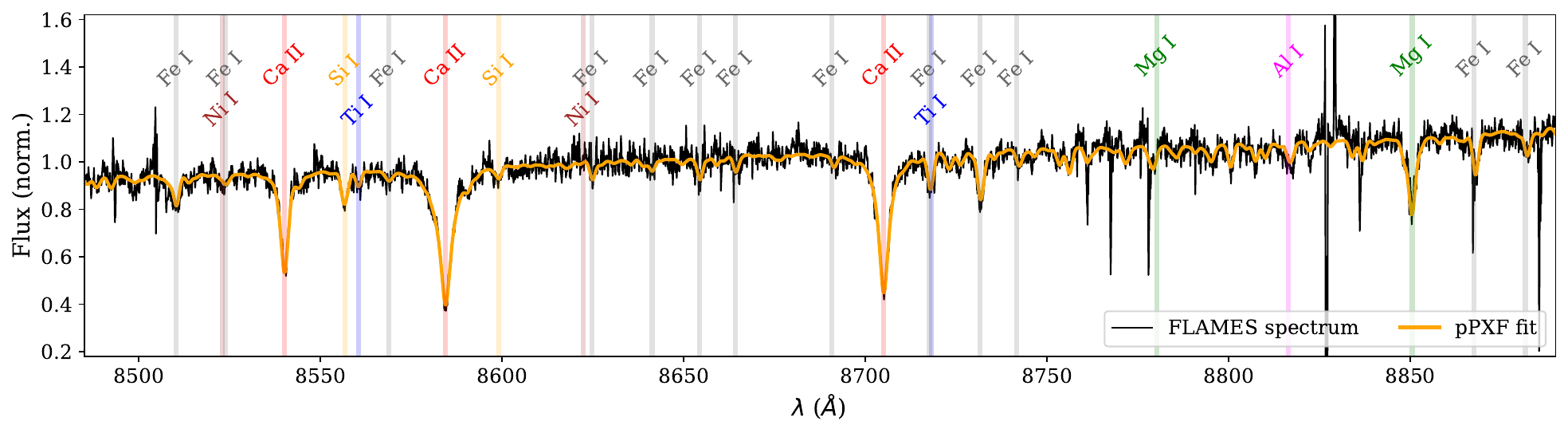}
    \caption{Example FLAMES GIRAFFE spectrum of a GC near NGC\,1404 (black). The orange line shows a \textsc{pPXF} fit. Prominent absorption lines are indicated by the vertical lines.}
    \label{fig:spec_with_lines}
\end{figure*}

\subsection{GC stellar mass function and black hole studies}
For the distance measurements, we shall derive velocity dispersions using full spectrum fitting of the FLAMES spectra with appropriate stellar and stellar population models. The dispersions can also be used to study the dynamical properties of GCs. Internal stellar velocity dispersions together with size measurements constrain the dynamical $M/L$ ratio of GCs \citep{Martini2004, Rejkuba2007, Strader2009}. Comparing the dynamical $M/L$ with expectations from the evolution of a simple stellar population offers insights into the present-day stellar mass functions of GCs as well as their past dynamical evolution (e.g. \citealt{Strader2011}). For example, expulsion from low-mass stars and expansion due to two-body relaxation alters the dynamical $M/L$, while the stellar population $M/L$ changes with the passive stellar evolution of the stars within the GC (e.g. \citealt{Vesperini1997,Takahashi2000,Baumgardt2023_MF}). 

Due to the lack of high-resolution spectroscopy to measure velocity dispersions in lower-mass GCs, only the dynamical $M/L$ ratios of UCDs have been studied so far. Many UCDs show an elevated dynamical $M/L$ compared to the stellar population $M/L$ (e.g. \citealt{Hasegan2005, Hilker2007, Mieske2008, Forbes2014, Janz2015}). Three potential causes are usually discussed: i) deviation from the canonical IMF, either in form of a top- or bottom-heavy IMF (e.g. \citealt{Mieske2008, Dabringhausen2010}), ii) dark matter (e.g. \citealt{BaumgardtMieske2008}), or iii) the presence of massive black holes (e.g. \citealt{Mieske2013}).
High-resolution integral field spectroscopy and detailed dynamical modelling of some UCDs has indeed found evidence for supermassive black holes in UCDs (e.g. \citealt{Seth2014, Afanasiev2018, Ahn2018, FerreMateu2021, Taylor2025}), which gives further evidence that at least some UCDs are the remnant nuclear star clusters of disrupted satellite galaxies (e.g. \citealt{Pfeffer2014, Wang2023}). 

In principle, the same argumentation can be used to explore the dynamical $M/L$ of GCs and UCDs that GOLD will provide; however, it is still unclear whether GCs are expected to host massive central black holes. Even in the best-studied MW GCs the presence of a central massive black hole remains controversial (e.g. \citealt{Tremou2018, Baumgardt2019}). Interestingly, perhaps the best evidence comes from $\omega$ Cen \citep{Haeberle2024}, which itself is likely a remnant nuclear star cluster rather than a GC per se. The GOLD survey includes many GCs and UCDs more massive than $\omega$ Cen, and the potential impact of any putative massive central black holes on their stellar velocity dispersions is yet to be explored.

\subsection{Chemical abundances from high-resolution spectra}
\label{sect:Chemistry}
Another science goal of GOLD is to make use of the chemical abundance information embedded in the CaT spectral region. While this spectral region has been used before as a metallicity indicator \citep{Usher2012, Usher2019_WAGGS}, the higher spectral resolution may allow us to go beyond average metallicities and instead study the abundances of individual elements in a similar fashion as has been done for GCs in the Local Group (e.g. \citealt{Larsen2022, Gvozdenko2022, Gvozdenko2024}).

In terms of chemical abundances, the CaT region has a number of chemical species relevant to a variety of scientific problems present in a relatively small wavelength range as illustrated in Fig. \ref{fig:spec_with_lines}. We explored which lines are measurable at our spectral resolution and S/N by comparison of stellar models, calculated with the Python code iSpy3 \citep{Larsen2020}, with the spectrum of the K1.5 III star Arcturus (\citealt{Hinkle2000}, see \citealt{Larsen2022} for details).  We  identified relatively strong lines of Ca, Mg, Ti, Si, Fe, and Al in this region, and then checked that these lines are present in the GOLD spectra with sufficient S/N ($>40$ / pix). In the future, we plan to use similar methods to the ones described in \cite{Larsen2022, Gvozdenko2022} or \cite{Gvozdenko2024} to measure abundances from the GOLD CaT spectra with uncertainties of $\sim 0.1$ dex.

In particular, the Al doublet at 8772.87$\AA$ and 8773.90$\AA$ potentially offers a clean measure of the Al abundance, since it is prominent in cool red giant branch stars, suffers minimal blending, and is only weakly affected by non-local thermodynamic equilibrium \citep{NordlanderLind2017}. The combination of Al and Mg, for example, therefore offers the possibility of exploring the effects of multiple populations in extragalactic GCs, along with potential discrimination between in situ and ex situ populations (e.g. \citealt{Hawkins2015, Horta2020}).

\subsection{Galaxy halo dynamics and clustering analysis} 
For many decades, GCs have been used as dynamical probes of the host galaxy gravitational potential (e.g. \citealt{Zepf2000, Cote2001, Strader2011b, Schuberth2006, Versic2024, Chaturvedi2026}). Since typical velocity dispersions of massive galaxies are of the order of 100 km s$^{-1}$ or more, the velocity uncertainties provided by lower-resolution spectroscopy ($\Delta v \sim 5-20$ km s$^{-1}$) are usually sufficient to constrain the enclosed mass. 

However, to explore potential dynamical substructures, higher-velocity fidelity is required. The typical velocity precision of GOLD of $\sim 1-2$ km s$^{-1}$ corresponds to the velocity dispersion of a typical dwarf galaxy with dynamical masses of $\sim10^{5-6}M_\odot$ (e.g. \citealt{McConnachie2012}). Thus, dynamical information combined with spatial and chemical information can plausibly be used to look for low-mass accreted substructures in massive galaxy haloes (e.g. \citealt{Woodley2011,Veljanoski2013, Callingham2022,Lewis2023}).

\section{Conclusions}
\label{sect:conclusions}
This paper introduces the GOLD project, an ongoing effort to collect the first large sample of higher resolution spectroscopy of extragalactic GCs. Using the FLAMES GIRAFFE spectrograph, we have so far collected CaT spectra for  17 nearby galaxies that span distances of between 2 and 20 Mpc. The GOLD target list includes galaxies with a range of masses, morphologies, environments, and star formation characteristics.

The FLAMES spectra cover a wavelength range from 8480 to 9000\AA\,at a resolution of $R \sim 18,000$. This high resolution not only allowed us to measure LOS mean velocities with small uncertainties ($\Delta v \sim 1 - 2$ km s$^{-1}$), but will also be used to measure velocity dispersions down to $\sigma \sim 3 - 4$ km s$^{-1}$, as shown in our pilot work \citep{Beasley2025, Fahrion2026a}. 

In this paper, we measured LOS mean velocities for 1701 sources and used the velocities to separate genuine extragalactic GCs and UCDs from MW foreground stars, background galaxies, and low-S/N sources. Our sample so far includes 1062 GCs and 101 UCDs. We make our velocity catalogues publicly available. Cross-matching with available catalogues from the literature, we found that 465 GCs and UCDs have $V$-band magnitudes available, 769 have $g$-band magnitudes, and 556 are also matched with Gaia photometry. Additionally, 239 GCs and 75 UCDs have available size measurements, and 423 GCs have previously been observed with spectrographs and have other velocity measurements available that largely agree with our measurements.

While the focus of this paper was to introduce the project and its data, we also discussed the science cases that we aim to address with GOLD. As will be presented in future work, the main goal of GOLD is to use the velocity dispersions to measure distances to our target galaxies using the correlation between internal velocity dispersions and absolute magnitudes for GCs. Additionally, we aim to use the large sample of velocity dispersions of extragalactic GCs to study the dynamical evolution of GCs, employ the information-rich CaT region to study chemical abundances, and use precision LOS mean velocities to potentially identify chemo-dynamical substructure in galaxy haloes.

\section{Data availability}
Tables \ref{tab:losv_catalogue} and \ref{tab:foreground_stars} are only available in electronic form at the CDS via anonymous ftp to \url{cdsarc.u-strasbg.fr} (130.79.128.5) or via \url{http://cdsweb.u-strasbg.fr/cgi-bin/qcat?J/A+A/}.

\begin{acknowledgements}
     We thank the referee for useful comments that have improved this manuscript. KF acknowledges funding from the European Union’s Horizon Europe research and innovation programme under the Marie Sk\l{}odowska-Curie grant agreement No 101103830. 
     MAB, AFM, IOC acknowledge funding from the Spanish Science Ministry via the "MATRYOSHKA" project PID2024-16088NB-I00. MAB acknowledges financial support from the Spanish Agencia Estatal de Investigación (AEI) through the Severo Ochoa Centre of Excellence accreditation awarded to the Instituto de Astrofísica de Canarias, grant CEX2025-001609-S, funded by MICIU/AEI/10.13039/501100011033. 
     IOC acknowledges support from PREP2024-003019. 
     AFM also acknowledges support from RYC2021-031099-I. 
     AG was supported by the STFC through consolidated grant ST/X001075/1. 
     O.M. is grateful to the Swiss National Science Foundation for financial support under the grant number PZ00P2\_202104. 
     DD acknowledges funding from the European Union’s Horizon Europe research and innovation programme under the Marie Sk\l{}odowska-Curie grant agreement No 101103062. 
     T.J. acknowledges the support from the MUNI Award in Science and Humanities (MUNI/I/1762/2023). 
     This work made use of Astropy: \footnote{https://www.astropy.org} a community-developed core Python package and an ecosystem of tools and resources for astronomy \citep{astropy2013, astropy2018, astropy2022}.
     Based on observations made with ESO Telescopes at the La Silla Paranal Observatory under programme IDs 114.274W.001, 115.27YL.001, and 116.28QP.001. 
\end{acknowledgements}

\bibliographystyle{aa} 
\bibliography{References} 

@article{Callingham2022,
    author = {Callingham, Thomas M and Cautun, Marius and Deason, Alis J and Frenk, Carlos S and Grand, Robert J J and Marinacci, Federico},
    title = {The chemo-dynamical groups of Galactic globular clusters},
    journal = {Monthly Notices of the Royal Astronomical Society},
    volume = {513},
    number = {3},
    pages = {4107-4129},
    year = {2022},
    month = {07},
    issn = {0035-8711},
    doi = {10.1093/mnras/stac1145},
    url = {https://doi.org/10.1093/mnras/stac1145},
    eprint = {https://academic.oup.com/mnras/article-pdf/513/3/4107/43722845/stac1145.pdf},
}

@article{Lewis2023,
    author = {Lewis, Geraint F and Brewer, Brendon J and Mackey, Dougal and Ferguson, Annette M N and Li, Yuan (Cher) and Adams, Tim},
    title = {Chemo-dynamical substructure in the M31 inner halo globular clusters: further evidence for a recent accretion event},
    journal = {Monthly Notices of the Royal Astronomical Society},
    volume = {518},
    number = {4},
    pages = {5778-5787},
    year = {2023},
    month = {02},
    issn = {0035-8711},
    doi = {10.1093/mnras/stac3325},
    url = {https://doi.org/10.1093/mnras/stac3325},
    eprint = {https://academic.oup.com/mnras/article-pdf/518/4/5778/47852400/stac3325.pdf},
}

@ARTICLE{Woodley2011,
       author = {{Woodley}, Kristin A. and {Harris}, William E.},
        title = "{Possible Subgroups of Globular Clusters and Planetary Nebulae in NGC 5128}",
      journal = {\aj},
         year = 2011,
        month = jan,
       volume = {141},
       number = {1},
          eid = {27},
        pages = {27},
          doi = {10.1088/0004-6256/141/1/27},
archivePrefix = {arXiv},
       eprint = {1011.3558},
 primaryClass = {astro-ph.CO},
       adsurl = {https://ui.adsabs.harvard.edu/abs/2011AJ....141...27W}
}

@ARTICLE{Takahashi2000,
       author = {{Takahashi}, K. and {Lee}, H.~M.},
        title = "{Evolution of multimass globular clusters in the Galactic tidal field with the effects of velocity anisotropy}",
      journal = {\mnras},
         year = 2000,
        month = aug,
       volume = {316},
       number = {3},
        pages = {671-683},
          doi = {10.1046/j.1365-8711.2000.03594.x},
archivePrefix = {arXiv},
       eprint = {astro-ph/9909006},
 primaryClass = {astro-ph},
       adsurl = {https://ui.adsabs.harvard.edu/abs/2000MNRAS.316..671T}
}

@ARTICLE{Vesperini1997,
       author = {{Vesperini}, E. and {Heggie}, D.~C.},
        title = "{On the effects of dynamical evolution on the initial mass function of globular clusters}",
      journal = {\mnras},
         year = 1997,
        month = aug,
       volume = {289},
       number = {4},
        pages = {898-920},
          doi = {10.1093/mnras/289.4.898},
archivePrefix = {arXiv},
       eprint = {astro-ph/9705073},
 primaryClass = {astro-ph},
       adsurl = {https://ui.adsabs.harvard.edu/abs/1997MNRAS.289..898V}
}

@ARTICLE{Baumgardt2023_MF,
       author = {{Baumgardt}, H. and {H{\'e}nault-Brunet}, V. and {Dickson}, N. and {Sollima}, A.},
        title = "{Evidence for a bottom-light initial mass function in massive star clusters}",
      journal = {\mnras},
         year = 2023,
        month = may,
       volume = {521},
       number = {3},
        pages = {3991-4008},
          doi = {10.1093/mnras/stad631},
archivePrefix = {arXiv},
       eprint = {2303.01636},
 primaryClass = {astro-ph.GA},
       adsurl = {https://ui.adsabs.harvard.edu/abs/2023MNRAS.521.3991B}
}

@ARTICLE{Beasley2000,
       author = {{Beasley}, M.~A. and {Sharples}, R.~M.},
        title = "{Spectroscopy of globular cluster candidates in the Sculptor group galaxies NGC 253 and 55}",
      journal = {\mnras},
         year = 2000,
        month = feb,
       volume = {311},
       number = {4},
        pages = {673-682},
          doi = {10.1046/j.1365-8711.2000.03080.x},
archivePrefix = {arXiv},
       eprint = {astro-ph/9909230},
 primaryClass = {astro-ph},
       adsurl = {https://ui.adsabs.harvard.edu/abs/2000MNRAS.311..673B}
}

@ARTICLE{Baumgardt2018,
       author = {{Baumgardt}, H. and {Hilker}, M.},
        title = "{A catalogue of masses, structural parameters, and velocity dispersion profiles of 112 Milky Way globular clusters}",
      journal = {\mnras},
         year = 2018,
        month = aug,
       volume = {478},
       number = {2},
        pages = {1520-1557},
          doi = {10.1093/mnras/sty1057},
archivePrefix = {arXiv},
       eprint = {1804.08359},
 primaryClass = {astro-ph.GA},
       adsurl = {https://ui.adsabs.harvard.edu/abs/2018MNRAS.478.1520B}
}

@ARTICLE{Monty2024,
       author = {{Monty}, Stephanie and {Belokurov}, Vasily and {Sanders}, Jason L. and {Hansen}, Terese T. and {Sakari}, Charli M. and {McKenzie}, Madeleine and {Myeong}, GyuChul and {Davies}, Elliot Y. and {Ardern-Arentsen}, Anke and {Massari}, Davide},
        title = "{The ratio of [Eu/{\ensuremath{\alpha}}] differentiates accreted/in situ Milky Way stars across metallicities, as indicated by both field stars and globular clusters}",
      journal = {\mnras},
         year = 2024,
        month = sep,
       volume = {533},
       number = {2},
        pages = {2420-2440},
          doi = {10.1093/mnras/stae1895},
archivePrefix = {arXiv},
       eprint = {2405.08963},
 primaryClass = {astro-ph.GA},
       adsurl = {https://ui.adsabs.harvard.edu/abs/2024MNRAS.533.2420M}
}

@ARTICLE{Anand2021,
       author = {{Anand}, Gagandeep S. and {Rizzi}, Luca and {Tully}, R. Brent and {Shaya}, Edward J. and {Karachentsev}, Igor D. and {Makarov}, Dmitry I. and {Makarova}, Lidia and {Wu}, Po-Feng and {Dolphin}, Andrew E. and {Kourkchi}, Ehsan},
        title = "{The Extragalactic Distance Database: The Color-Magnitude Diagrams/Tip of the Red Giant Branch Distance Catalog}",
      journal = {\aj},
         year = 2021,
        month = aug,
       volume = {162},
       number = {2},
          eid = {80},
        pages = {80},
          doi = {10.3847/1538-3881/ac0440},
archivePrefix = {arXiv},
       eprint = {2104.02649},
 primaryClass = {astro-ph.GA},
       adsurl = {https://ui.adsabs.harvard.edu/abs/2021AJ....162...80A}
}

@ARTICLE{McQuinn2016,
       author = {{McQuinn}, Kristen. B.~W. and {Skillman}, Evan D. and {Dolphin}, Andrew E. and {Berg}, Danielle and {Kennicutt}, Robert},
        title = "{The Distance to M104}",
      journal = {\aj},
         year = 2016,
        month = nov,
       volume = {152},
       number = {5},
          eid = {144},
        pages = {144},
          doi = {10.3847/0004-6256/152/5/144},
       adsurl = {https://ui.adsabs.harvard.edu/abs/2016AJ....152..144M}
}

@INCOLLECTION{Beasley2020,
       author = {{Beasley}, Michael A.},
        title = "{Globular Cluster Systems and Galaxy Formation}",
    booktitle = {Reviews in Frontiers of Modern Astrophysics; From Space Debris to Cosmology},
         year = 2020,
       editor = {{Kab{\'a}th}, Petr and {Jones}, David and {Skarka}, Marek},
        pages = {245-277},
          doi = {10.1007/978-3-030-38509-5_9},
       adsurl = {https://ui.adsabs.harvard.edu/abs/2020rfma.book..245B}
}

@ARTICLE{BrodieStrader2006,
       author = {{Brodie}, Jean P. and {Strader}, Jay},
        title = "{Extragalactic Globular Clusters and Galaxy Formation}",
      journal = {\araa},
         year = 2006,
        month = sep,
       volume = {44},
       number = {1},
        pages = {193-267},
          doi = {10.1146/annurev.astro.44.051905.092441},
archivePrefix = {arXiv},
       eprint = {astro-ph/0602601},
 primaryClass = {astro-ph},
       adsurl = {https://ui.adsabs.harvard.edu/abs/2006ARA&A..44..193B}
}

@ARTICLE{Rhode2001,
       author = {{Rhode}, Katherine L. and {Zepf}, Stephen E.},
        title = "{The Globular Cluster System in the Outer Regions of NGC 4472}",
      journal = {\aj},
         year = 2001,
        month = jan,
       volume = {121},
       number = {1},
        pages = {210-224},
          doi = {10.1086/318039},
archivePrefix = {arXiv},
       eprint = {astro-ph/0010065},
 primaryClass = {astro-ph},
       adsurl = {https://ui.adsabs.harvard.edu/abs/2001AJ....121..210R}
}

@ARTICLE{Globsi,
       author = {{Dold}, Dominik and {Fahrion}, Katja},
        title = "{Evaluating the feasibility of interpretable machine learning for globular cluster detection}",
      journal = {\aap},
         year = 2022,
        month = jul,
       volume = {663},
          eid = {A81},
        pages = {A81},
          doi = {10.1051/0004-6361/202243354},
archivePrefix = {arXiv},
       eprint = {2204.00017},
 primaryClass = {astro-ph.GA},
       adsurl = {https://ui.adsabs.harvard.edu/abs/2022A&A...663A..81D}
}

@ARTICLE{Shanks2013,
       author = {{Shanks}, T. and {Belokurov}, V. and {Chehade}, B. and {Croom}, S.~M. and {Findlay}, J.~R. and {Gonzalez-Solares}, E. and {Irwin}, M.~J. and {Koposov}, S. and {Mann}, R.~G. and {Metcalfe}, N. and {Murphy}, D.~N.~A. and {Norberg}, P.~R. and {Read}, M.~A. and {Sutorius}, E. and {Worseck}, G.},
        title = "{VST ATLAS First Science Results}",
      journal = {The Messenger},
         year = 2013,
        month = dec,
       volume = {154},
        pages = {38-40},
       adsurl = {https://ui.adsabs.harvard.edu/abs/2013Msngr.154...38S}
}

@ARTICLE{Cantiello2020,
       author = {{Cantiello}, Michele and {Venhola}, Aku and {Grado}, Aniello and {Paolillo}, Maurizio and {D'Abrusco}, Raffaele and {Raimondo}, Gabriella and {Quintini}, Massimo and {Hilker}, Michael and {Mieske}, Steffen and {Tortora}, Crescenzo and {Spavone}, Marilena and {Capaccioli}, Massimo and {Iodice}, Enrica and {Peletier}, Reynier and {Barroso}, Jes{\'u}s Falc{\'o}n and {Limatola}, Luca and {Napolitano}, Nicola and {Schipani}, Pietro and {van de Ven}, Glenn and {Gentile}, Fabrizio and {Covone}, Giovanni},
        title = "{The Fornax Deep Survey with VST. IX. Catalog of sources in the FDS area with an example study for globular clusters and background galaxies}",
      journal = {\aap},
         year = 2020,
        month = jul,
       volume = {639},
          eid = {A136},
        pages = {A136},
          doi = {10.1051/0004-6361/202038137},
archivePrefix = {arXiv},
       eprint = {2005.12085},
 primaryClass = {astro-ph.GA},
       adsurl = {https://ui.adsabs.harvard.edu/abs/2020A&A...639A.136C}
}

@ARTICLE{ZepfAshman1993,
       author = {{Zepf}, S.~E. and {Ashman}, K.~M.},
        title = "{Globular cluster systems formed in galaxy mergers.}",
      journal = {\mnras},
         year = 1993,
        month = oct,
       volume = {264},
        pages = {611-618},
          doi = {10.1093/mnras/264.3.611},
       adsurl = {https://ui.adsabs.harvard.edu/abs/1993MNRAS.264..611Z}
}

@ARTICLE{Vazdekis2003,
       author = {{Vazdekis}, A. and {Cenarro}, A.~J. and {Gorgas}, J. and {Cardiel}, N. and {Peletier}, R.~F.},
        title = "{Empirical calibration of the near-infrared CaII triplet - IV. The stellar population synthesis models}",
      journal = {\mnras},
         year = 2003,
        month = apr,
       volume = {340},
       number = {4},
        pages = {1317-1345},
          doi = {10.1046/j.1365-8711.2003.06413.x},
archivePrefix = {arXiv},
       eprint = {astro-ph/0303297},
 primaryClass = {astro-ph},
       adsurl = {https://ui.adsabs.harvard.edu/abs/2003MNRAS.340.1317V}
}

@ARTICLE{Coelho2014,
       author = {{Coelho}, P.~R.~T.},
        title = "{A new library of theoretical stellar spectra with scaled-solar and {\ensuremath{\alpha}}-enhanced mixtures}",
      journal = {\mnras},
         year = 2014,
        month = may,
       volume = {440},
       number = {2},
        pages = {1027-1043},
          doi = {10.1093/mnras/stu365},
archivePrefix = {arXiv},
       eprint = {1404.3243},
 primaryClass = {astro-ph.SR},
       adsurl = {https://ui.adsabs.harvard.edu/abs/2014MNRAS.440.1027C}
}

@ARTICLE{Cenarro2004,
       author = {{Cenarro}, A.~J. and {S{\'a}nchez-Bl{\'a}zquez}, P. and {Cardiel}, N. and {Gorgas}, J.},
        title = "{Early-Type Galaxies in the Coma Cluster: A New Piece in the Calcium Puzzle}",
      journal = {\apjl},
         year = 2004,
        month = oct,
       volume = {614},
       number = {2},
        pages = {L101-L104},
          doi = {10.1086/425862},
archivePrefix = {arXiv},
       eprint = {astro-ph/0409301},
 primaryClass = {astro-ph},
       adsurl = {https://ui.adsabs.harvard.edu/abs/2004ApJ...614L.101C}
}

@ARTICLE{Foster2011,
       author = {{Foster}, Caroline and {Spitler}, Lee R. and {Romanowsky}, Aaron J. and {Forbes}, Duncan A. and {Pota}, Vincenzo and {Bekki}, Kenji and {Strader}, Jay and {Proctor}, Robert N. and {Arnold}, Jacob A. and {Brodie}, Jean P.},
        title = "{Global properties of 'ordinary' early-type galaxies: photometry and spectroscopy of stars and globular clusters in NGC 4494}",
      journal = {\mnras},
         year = 2011,
        month = aug,
       volume = {415},
       number = {4},
        pages = {3393-3416},
          doi = {10.1111/j.1365-2966.2011.18965.x},
archivePrefix = {arXiv},
       eprint = {1104.5503},
 primaryClass = {astro-ph.CO},
       adsurl = {https://ui.adsabs.harvard.edu/abs/2011MNRAS.415.3393F}
}

@ARTICLE{Foster2010,
       author = {{Foster}, Caroline and {Forbes}, Duncan A. and {Proctor}, Robert N. and {Strader}, Jay and {Brodie}, Jean P. and {Spitler}, Lee R.},
        title = "{Deriving Metallicities from the Integrated Spectra of Extragalactic Globular Clusters Using the Near-infrared Calcium Triplet}",
      journal = {\aj},
         year = 2010,
        month = apr,
       volume = {139},
       number = {4},
        pages = {1566-1578},
          doi = {10.1088/0004-6256/139/4/1566},
archivePrefix = {arXiv},
       eprint = {1002.1107},
 primaryClass = {astro-ph.CO},
       adsurl = {https://ui.adsabs.harvard.edu/abs/2010AJ....139.1566F}
}

@ARTICLE{Sakari2022,
       author = {{Sakari}, Charli M. and {Wallerstein}, George},
        title = "{Metallicities of outer halo M31 globular clusters from integrated light calcium-II triplet spectroscopy}",
      journal = {\mnras},
         year = 2022,
        month = jun,
       volume = {512},
       number = {4},
        pages = {4819-4834},
          doi = {10.1093/mnras/stac752},
archivePrefix = {arXiv},
       eprint = {2203.08840},
 primaryClass = {astro-ph.GA},
       adsurl = {https://ui.adsabs.harvard.edu/abs/2022MNRAS.512.4819S}
}

@ARTICLE{Dumont2022,
       author = {{Dumont}, Antoine and {Seth}, Anil C. and {Strader}, Jay and {Voggel}, Karina and {Sand}, David J. and {Hughes}, Allison K. and {Caldwell}, Nelson and {Crnojevi{\'c}}, Denija and {Mateo}, Mario and {Bailey}, John I. and {Forbes}, Duncan A.},
        title = "{A Population of Luminous Globular Clusters and Stripped Nuclei with Elevated Mass to Light Ratios around NGC 5128}",
      journal = {\apj},
         year = 2022,
        month = apr,
       volume = {929},
       number = {2},
          eid = {147},
        pages = {147},
          doi = {10.3847/1538-4357/ac551c},
archivePrefix = {arXiv},
       eprint = {2112.04504},
 primaryClass = {astro-ph.GA},
       adsurl = {https://ui.adsabs.harvard.edu/abs/2022ApJ...929..147D}
}

@ARTICLE{Strader2011,
       author = {{Strader}, Jay and {Caldwell}, Nelson and {Seth}, Anil C.},
        title = "{Star Clusters in M31. V. Internal Dynamical Trends: Some Troublesome, Some Reassuring}",
      journal = {\aj},
         year = 2011,
        month = jul,
       volume = {142},
       number = {1},
          eid = {8},
        pages = {8},
          doi = {10.1088/0004-6256/142/1/8},
archivePrefix = {arXiv},
       eprint = {1104.4649},
 primaryClass = {astro-ph.CO},
       adsurl = {https://ui.adsabs.harvard.edu/abs/2011AJ....142....8S}
}

@ARTICLE{Sakari2016,
       author = {{Sakari}, Charli M. and {Shetrone}, Matthew D. and {Schiavon}, Ricardo P. and {Bizyaev}, Dmitry and {Allende Prieto}, Carlos and {Beers}, Timothy C. and {Caldwell}, Nelson and {Garc{\'\i}a-Hern{\'a}ndez}, D.~A. and {Lucatello}, Sara and {Majewski}, Steven and {O'Connell}, Robert W. and {Pan}, Kaike and {Strader}, Jay},
        title = "{Infrared High-resolution Integrated Light Spectral Analyses of M31 Globular Clusters from APOGEE}",
      journal = {\apj},
         year = 2016,
        month = oct,
       volume = {829},
       number = {2},
          eid = {116},
        pages = {116},
          doi = {10.3847/0004-637X/829/2/116},
archivePrefix = {arXiv},
       eprint = {1607.06811},
 primaryClass = {astro-ph.GA},
       adsurl = {https://ui.adsabs.harvard.edu/abs/2016ApJ...829..116S}
}

@ARTICLE{Colucci2017,
       author = {{Colucci}, Janet E. and {Bernstein}, Rebecca A. and {McWilliam}, Andrew},
        title = "{Globular Cluster Abundances from High-resolution, Integrated-light Spectroscopy. II. Expanding the Metallicity Range for Old Clusters and Updated Analysis Techniques}",
      journal = {\apj},
         year = 2017,
        month = jan,
       volume = {834},
       number = {2},
          eid = {105},
        pages = {105},
          doi = {10.3847/1538-4357/834/2/105},
archivePrefix = {arXiv},
       eprint = {1611.02734},
 primaryClass = {astro-ph.GA},
       adsurl = {https://ui.adsabs.harvard.edu/abs/2017ApJ...834..105C}
}

@ARTICLE{CabreraZiri2022,
       author = {{Cabrera-Ziri}, Ivan and {Conroy}, Charlie},
        title = "{Deriving ages and horizontal branch properties of integrated stellar populations}",
      journal = {\mnras},
         year = 2022,
        month = mar,
       volume = {511},
       number = {1},
        pages = {341-355},
          doi = {10.1093/mnras/stac012},
archivePrefix = {arXiv},
       eprint = {2112.07562},
 primaryClass = {astro-ph.GA},
       adsurl = {https://ui.adsabs.harvard.edu/abs/2022MNRAS.511..341C}
}

@ARTICLE{Gvozdenko2026,
       author = {{Gvozdenko}, A. and {Beasley}, M.~A. and {Fahrion}, K. and {Larsen}, S.~S. and {La Barbara}, F. and {Vazdekis}, A. and {Mart\'{\i}n-Navarro}, I},
        title = "{On the feasibility of using $\text{[Eu/}\alpha\text{]}$ to identify accreted and in-situ extragalactic globular clusters}",
      journal = {\mnras},
         year = 2026,
         volume = {submitted},
}

@ARTICLE{Gvozdenko2024,
       author = {{Gvozdenko}, A. and {Larsen}, S.~S. and {Beasley}, M.~A. and {Cabrera-Ziri}, I. and {Eitner}, P. and {Battaglia}, G. and {Leaman}, R.},
        title = "{Detailed chemical composition of the globular cluster Sextans A GC-1 on the outskirts of the Local Group}",
      journal = {\aap},
         year = 2024,
        month = may,
       volume = {685},
          eid = {A154},
        pages = {A154},
          doi = {10.1051/0004-6361/202346859},
archivePrefix = {arXiv},
       eprint = {2403.10597},
 primaryClass = {astro-ph.GA},
       adsurl = {https://ui.adsabs.harvard.edu/abs/2024A&A...685A.154G}
}

@ARTICLE{Gvozdenko2022,
       author = {{Gvozdenko}, A. and {Larsen}, S.~S. and {Beasley}, M.~A. and {Brodie}, J.},
        title = "{Chemical composition of the young massive cluster NGC 1569-B}",
      journal = {\aap},
         year = 2022,
        month = oct,
       volume = {666},
          eid = {A159},
        pages = {A159},
          doi = {10.1051/0004-6361/202243415},
archivePrefix = {arXiv},
       eprint = {2209.11779},
 primaryClass = {astro-ph.GA},
       adsurl = {https://ui.adsabs.harvard.edu/abs/2022A&A...666A.159G}
}

@ARTICLE{Larsen2022,
       author = {{Larsen}, S.~S. and {Eitner}, P. and {Magg}, E. and {Bergemann}, M. and {Moltzer}, C.~A.~S. and {Brodie}, J.~P. and {Romanowsky}, A.~J. and {Strader}, J.},
        title = "{The chemical composition of globular clusters in the Local Group}",
      journal = {\aap},
         year = 2022,
        month = apr,
       volume = {660},
          eid = {A88},
        pages = {A88},
          doi = {10.1051/0004-6361/202142243},
archivePrefix = {arXiv},
       eprint = {2112.00081},
 primaryClass = {astro-ph.GA},
       adsurl = {https://ui.adsabs.harvard.edu/abs/2022A&A...660A..88L}
}

@ARTICLE{Brodie2014,
       author = {{Brodie}, Jean P. and {Romanowsky}, Aaron J. and {Strader}, Jay and {Forbes}, Duncan A. and {Foster}, Caroline and {Jennings}, Zachary G. and {Pastorello}, Nicola and {Pota}, Vincenzo and {Usher}, Christopher and {Blom}, Christina and {Kader}, Justin and {Roediger}, Joel C. and {Spitler}, Lee R. and {Villaume}, Alexa and {Arnold}, Jacob A. and {Kartha}, Sreeja S. and {Woodley}, Kristin A.},
        title = "{The SAGES Legacy Unifying Globulars and GalaxieS Survey (SLUGGS): Sample Definition, Methods, and Initial Results}",
      journal = {\apj},
         year = 2014,
        month = nov,
       volume = {796},
       number = {1},
          eid = {52},
        pages = {52},
          doi = {10.1088/0004-637X/796/1/52},
archivePrefix = {arXiv},
       eprint = {1405.2079},
 primaryClass = {astro-ph.GA},
       adsurl = {https://ui.adsabs.harvard.edu/abs/2014ApJ...796...52B}
}

@ARTICLE{Villaume2019,
       author = {{Villaume}, Alexa and {Romanowsky}, Aaron J. and {Brodie}, Jean and {Strader}, Jay},
        title = "{New Constraints on Early-type Galaxy Assembly from Spectroscopic Metallicities of Globular Clusters in M87}",
      journal = {\apj},
         year = 2019,
        month = jul,
       volume = {879},
       number = {1},
          eid = {45},
        pages = {45},
          doi = {10.3847/1538-4357/ab24d7},
archivePrefix = {arXiv},
       eprint = {1905.12641},
 primaryClass = {astro-ph.GA},
       adsurl = {https://ui.adsabs.harvard.edu/abs/2019ApJ...879...45V}
}

@ARTICLE{Usher2012,
       author = {{Usher}, Christopher and {Forbes}, Duncan A. and {Brodie}, Jean P. and {Foster}, Caroline and {Spitler}, Lee R. and {Arnold}, Jacob A. and {Romanowsky}, Aaron J. and {Strader}, Jay and {Pota}, Vincenzo},
        title = "{The SLUGGS survey: calcium triplet-based spectroscopic metallicities for over 900 globular clusters}",
      journal = {\mnras},
         year = 2012,
        month = oct,
       volume = {426},
       number = {2},
        pages = {1475-1495},
          doi = {10.1111/j.1365-2966.2012.21801.x},
archivePrefix = {arXiv},
       eprint = {1207.6402},
 primaryClass = {astro-ph.CO},
       adsurl = {https://ui.adsabs.harvard.edu/abs/2012MNRAS.426.1475U}
}

@ARTICLE{Lagarde2017,
       author = {{Lagarde}, N. and {Robin}, A.~C. and {Reyl{\'e}}, C. and {Nasello}, G.},
        title = "{Population synthesis to constrain Galactic and stellar physics. I. Determining age and mass of thin-disc red-giant stars}",
      journal = {\aap},
         year = 2017,
        month = may,
       volume = {601},
          eid = {A27},
        pages = {A27},
          doi = {10.1051/0004-6361/201630253},
archivePrefix = {arXiv},
       eprint = {1702.01769},
 primaryClass = {astro-ph.SR},
       adsurl = {https://ui.adsabs.harvard.edu/abs/2017A&A...601A..27L}
}

@ARTICLE{Robin2003,
       author = {{Robin}, A.~C. and {Reyl{\'e}}, C. and {Derri{\`e}re}, S. and {Picaud}, S.},
        title = "{A synthetic view on structure and evolution of the Milky Way}",
      journal = {\aap},
         year = 2003,
        month = oct,
       volume = {409},
        pages = {523-540},
          doi = {10.1051/0004-6361:20031117},
       adsurl = {https://ui.adsabs.harvard.edu/abs/2003A&A...409..523R}
}

@ARTICLE{Czekaj2014,
       author = {{Czekaj}, M.~A. and {Robin}, A.~C. and {Figueras}, F. and {Luri}, X. and {Haywood}, M.},
        title = "{The Besan{\c{c}}on Galaxy model renewed. I. Constraints on the local star formation history from Tycho data}",
      journal = {\aap},
         year = 2014,
        month = apr,
       volume = {564},
          eid = {A102},
        pages = {A102},
          doi = {10.1051/0004-6361/201322139},
archivePrefix = {arXiv},
       eprint = {1402.3257},
 primaryClass = {astro-ph.GA},
       adsurl = {https://ui.adsabs.harvard.edu/abs/2014A&A...564A.102C}
}

@ARTICLE{Usher2019_WAGGS,
       author = {{Usher}, Christopher and {Beckwith}, Thomas and {Bellstedt}, Sabine and {Alabi}, Adebusola and {Chevalier}, Leonie and {Pastorello}, Nicola and {Cerulo}, Pierluigi and {Dalgleish}, Hannah S. and {Fraser-McKelvie}, Amelia and {Kamann}, Sebastian and {Penny}, Samantha and {Foster}, Caroline and {McDermid}, Richard and {Schiavon}, Ricardo P. and {Villaume}, Alexa},
        title = "{The WAGGS project - II. The reliability of the calcium triplet as a metallicity indicator in integrated stellar light}",
      journal = {\mnras},
         year = 2019,
        month = jan,
       volume = {482},
       number = {1},
        pages = {1275-1303},
          doi = {10.1093/mnras/sty2611},
archivePrefix = {arXiv},
       eprint = {1809.07650},
 primaryClass = {astro-ph.GA},
       adsurl = {https://ui.adsabs.harvard.edu/abs/2019MNRAS.482.1275U}
}

@ARTICLE{Usher2019,
       author = {{Usher}, Christopher and {Brodie}, Jean P. and {Forbes}, Duncan A. and {Romanowsky}, Aaron J. and {Strader}, Jay and {Pfeffer}, Joel and {Bastian}, Nate},
        title = "{The SLUGGS survey: measuring globular cluster ages using both photometry and spectroscopy}",
      journal = {\mnras},
         year = 2019,
        month = nov,
       volume = {490},
       number = {1},
        pages = {491-501},
          doi = {10.1093/mnras/stz2596},
archivePrefix = {arXiv},
       eprint = {1909.05753},
 primaryClass = {astro-ph.GA},
       adsurl = {https://ui.adsabs.harvard.edu/abs/2019MNRAS.490..491U}
}

@ARTICLE{Harris2023,
       author = {{Harris}, William E.},
        title = "{A Photometric Survey of Globular Cluster Systems in Brightest Cluster Galaxies}",
      journal = {\apjs},
         year = 2023,
        month = mar,
       volume = {265},
       number = {1},
          eid = {9},
        pages = {9},
          doi = {10.3847/1538-4365/acab5c},
archivePrefix = {arXiv},
       eprint = {2212.06174},
 primaryClass = {astro-ph.GA},
       adsurl = {https://ui.adsabs.harvard.edu/abs/2023ApJS..265....9H}
}

@ARTICLE{LomeliNunez2025,
       author = {{Lomel{\'\i}-N{\'u}{\~n}ez}, Luis and {Cortesi}, A. and {Smith Castelli}, A.~V. and {Buzzo}, M.~L. and {Mayya}, Y.~D. and {Fragkou}, Vasiliki and {Alzate-Trujillo}, J.~A. and {Haack}, R.~F. and {Calder{\'o}n}, J.~P. and {Lopes}, A.~R. and {Hilker}, Michael and {Grossi}, M. and {Men{\'e}ndez-Delmestre}, Kar{\'\i}n and {Gon{\c{c}}alves}, Thiago S. and {Chies-Santos}, Ana L. and {Guti{\'e}rrez-Soto}, L.~A. and {Lima-Dias}, Ciria and {Werner}, S.~V. and {Humire}, Pedro K. and {Thom de Souza}, R.~C. and {Alvarez-Candal}, A. and {Panda}, Swayamtrupta and {Chaturvedi}, Avinash and {Telles}, E. and {Mendes de Oliveira}, C. and {Kanaan}, A. and {Ribeiro}, T. and {Schoenell}, W.},
        title = "{The S-PLUS Fornax Project (S+FP): Mapping Globular Clusters Systems within 5 Virial Radii around NGC 1399}",
      journal = {\aj},
         year = 2025,
        month = may,
       volume = {169},
       number = {5},
          eid = {263},
        pages = {263},
          doi = {10.3847/1538-3881/adbf0c},
archivePrefix = {arXiv},
       eprint = {2503.15657},
 primaryClass = {astro-ph.GA},
       adsurl = {https://ui.adsabs.harvard.edu/abs/2025AJ....169..263L}
}

@ARTICLE{Chaturvedi2022,
       author = {{Chaturvedi}, Avinash and {Hilker}, Michael and {Cantiello}, Michele and {Napolitano}, Nicola R. and {van de Ven}, Glenn and {Spiniello}, Chiara and {Fahrion}, Katja and {Paolillo}, Maurizio and {Gatto}, Massimiliano and {Puzia}, Thomas},
        title = "{The Fornax Cluster VLT Spectroscopic Survey. III. Kinematical characterisation of globular clusters in the Fornax galaxy cluster}",
      journal = {\aap},
         year = 2022,
        month = jan,
       volume = {657},
          eid = {A93},
        pages = {A93},
          doi = {10.1051/0004-6361/202141334},
archivePrefix = {arXiv},
       eprint = {2109.08694},
 primaryClass = {astro-ph.GA},
       adsurl = {https://ui.adsabs.harvard.edu/abs/2022A&A...657A..93C}
}

@ARTICLE{Jordan2015,
       author = {{Jord{\'a}n}, Andr{\'e}s and {Peng}, Eric W. and {Blakeslee}, John P. and {C{\^o}t{\'e}}, Patrick and {Eyheramendy}, Susana and {Ferrarese}, Laura},
        title = "{The ACS Fornax Cluster Survey. XI. Catalog of Globular Cluster Candidates}",
      journal = {\apjs},
         year = 2015,
        month = nov,
       volume = {221},
       number = {1},
          eid = {13},
        pages = {13},
          doi = {10.1088/0067-0049/221/1/13},
       adsurl = {https://ui.adsabs.harvard.edu/abs/2015ApJS..221...13J}
}

@ARTICLE{Schuberth2006,
       author = {{Schuberth}, Y. and {Richtler}, T. and {Dirsch}, B. and {Hilker}, M. and {Larsen}, S.~S. and {Kissler-Patig}, M. and {Mebold}, U.},
        title = "{Dynamics of the NGC 4636 globular cluster system. An extremely dark matter dominated galaxy?}",
      journal = {\aap},
         year = 2006,
        month = nov,
       volume = {459},
       number = {2},
        pages = {391-406},
          doi = {10.1051/0004-6361:20053134},
archivePrefix = {arXiv},
       eprint = {astro-ph/0604309},
 primaryClass = {astro-ph},
       adsurl = {https://ui.adsabs.harvard.edu/abs/2006A&A...459..391S}
}

@ARTICLE{Dirsch2005,
       author = {{Dirsch}, B. and {Schuberth}, Y. and {Richtler}, T.},
        title = "{A wide-field photometric study of the globular cluster system of NGC 4636}",
      journal = {\aap},
         year = 2005,
        month = apr,
       volume = {433},
       number = {1},
        pages = {43-56},
          doi = {10.1051/0004-6361:20035737},
archivePrefix = {arXiv},
       eprint = {astro-ph/0503283},
 primaryClass = {astro-ph},
       adsurl = {https://ui.adsabs.harvard.edu/abs/2005A&A...433...43D}
}

@ARTICLE{Park2010,
       author = {{Park}, Hong Soo and {Lee}, Myung Gyoon and {Hwang}, Ho Seong and {Arimoto}, Nobuo and {Tamura}, Naoyuki and {Onodera}, Masato},
        title = "{The Globular Cluster System of the Virgo Giant Elliptical Galaxy NGC 4636. I. Subaru/Faint Object Camera and Spectrograph Spectroscopy and Database}",
      journal = {\apj},
         year = 2010,
        month = jan,
       volume = {709},
       number = {1},
        pages = {377-385},
          doi = {10.1088/0004-637X/709/1/377},
archivePrefix = {arXiv},
       eprint = {0912.0340},
 primaryClass = {astro-ph.CO},
       adsurl = {https://ui.adsabs.harvard.edu/abs/2010ApJ...709..377P}
}

@ARTICLE{ChiesSantos2011,
       author = {{Chies-Santos}, A.~L. and {Larsen}, S.~S. and {Wehner}, E.~M. and {Kuntschner}, H. and {Strader}, J. and {Brodie}, J.~P.},
        title = "{An optical/NIR survey of globular clusters in early-type galaxies. I. Introduction and data reduction procedures}",
      journal = {\aap},
         year = 2011,
        month = jan,
       volume = {525},
          eid = {A19},
        pages = {A19},
          doi = {10.1051/0004-6361/201015681},
archivePrefix = {arXiv},
       eprint = {1010.0687},
 primaryClass = {astro-ph.CO},
       adsurl = {https://ui.adsabs.harvard.edu/abs/2011A&A...525A..19C}
}

@ARTICLE{Oldham2015,
       author = {{Oldham}, L.~J. and {Auger}, M.~W.},
        title = "{Galaxy structure from multiple tracers - I. A census of M87's globular cluster populations}",
      journal = {\mnras},
         year = 2016,
        month = jan,
       volume = {455},
       number = {1},
        pages = {820-830},
          doi = {10.1093/mnras/stv2244},
archivePrefix = {arXiv},
       eprint = {1511.07046},
 primaryClass = {astro-ph.GA},
       adsurl = {https://ui.adsabs.harvard.edu/abs/2016MNRAS.455..820O}
}

@ARTICLE{Pota2015,
       author = {{Pota}, Vincenzo and {Brodie}, Jean P. and {Bridges}, Terry and {Strader}, Jay and {Romanowsky}, Aaron J. and {Villaume}, Alexa and {Jennings}, Zach and {Faifer}, Favio R. and {Pastorello}, Nicola and {Forbes}, Duncan A. and {Campbell}, Ainsley and {Usher}, Christopher and {Foster}, Caroline and {Spitler}, Lee R. and {Caldwell}, Nelson and {Forte}, Juan C. and {Norris}, Mark A. and {Zepf}, Stephen E. and {Beasley}, Michael A. and {Gebhardt}, Karl and {Hanes}, David A. and {Sharples}, Ray M. and {Arnold}, Jacob A.},
        title = "{A SLUGGS and Gemini/GMOS combined study of the elliptical galaxy M60: wide-field photometry and kinematics of the globular cluster system}",
      journal = {\mnras},
         year = 2015,
        month = jun,
       volume = {450},
       number = {2},
        pages = {1962-1983},
          doi = {10.1093/mnras/stv677},
archivePrefix = {arXiv},
       eprint = {1503.07520},
 primaryClass = {astro-ph.GA},
       adsurl = {https://ui.adsabs.harvard.edu/abs/2015MNRAS.450.1962P}
}

@ARTICLE{Liu2020,
       author = {{Liu}, Chengze and {C{\^o}t{\'e}}, Patrick and {Peng}, Eric W. and {Roediger}, Joel and {Zhang}, Hongxin and {Ferrarese}, Laura and {S{\'a}nchez-Janssen}, Ruben and {Guhathakurta}, Puragra and {Yang}, Xiaohu and {Jing}, Yipeng and {Alamo-Mart{\'\i}nez}, Karla and {Blakeslee}, John P. and {Boselli}, Alessandro and {Cuilandre}, Jean-Charles and {Duc}, Pierre-Alain and {Durrell}, Patrick and {Gwyn}, Stephen and {Jord{\'a}n}, Andres and {Ko}, Youkyung and {Lan{\c{c}}on}, Ariane and {Lim}, Sungsoon and {Longobardi}, Alessia and {Mei}, Simona and {Mihos}, J. Christopher and {Mu{\~n}oz}, Roberto and {Powalka}, Mathieu and {Puzia}, Thomas and {Spengler}, Chelsea and {Toloba}, Elisa},
        title = "{The Next Generation Virgo Cluster Survey. XXXIV. Ultracompact Dwarf Galaxies in the Virgo Cluster}",
      journal = {\apjs},
         year = 2020,
        month = sep,
       volume = {250},
       number = {1},
          eid = {17},
        pages = {17},
          doi = {10.3847/1538-4365/abad91},
archivePrefix = {arXiv},
       eprint = {2007.15275},
 primaryClass = {astro-ph.GA},
       adsurl = {https://ui.adsabs.harvard.edu/abs/2020ApJS..250...17L}
}

@ARTICLE{Forbes2017,
       author = {{Forbes}, Duncan A. and {Alabi}, Adebusola and {Brodie}, Jean P. and {Romanowsky}, Aaron J. and {Strader}, Jay and {Foster}, Caroline and {Usher}, Christopher and {Spitler}, Lee and {Bellstedt}, Sabine and {Pastorello}, Nicola and {Villaume}, Alexa and {Wasserman}, Asher and {Pota}, Vincenzo},
        title = "{The SLUGGS Survey: A Catalog of Over 4000 Globular Cluster Radial Velocities in 27 Nearby Early-type Galaxies}",
      journal = {\aj},
         year = 2017,
        month = mar,
       volume = {153},
       number = {3},
          eid = {114},
        pages = {114},
          doi = {10.3847/1538-3881/153/3/114},
archivePrefix = {arXiv},
       eprint = {1701.04835},
 primaryClass = {astro-ph.GA},
       adsurl = {https://ui.adsabs.harvard.edu/abs/2017AJ....153..114F}
}

@ARTICLE{Strader2012,
       author = {{Strader}, Jay and {Fabbiano}, Giuseppina and {Luo}, Bin and {Kim}, Dong-Woo and {Brodie}, Jean P. and {Fragos}, Tassos and {Gallagher}, John S. and {Kalogera}, Vassiliki and {King}, Andrew and {Zezas}, Andreas},
        title = "{Deep Chandra Monitoring Observations of NGC 4649. II. Wide-field Hubble Space Telescope Imaging of the Globular Clusters}",
      journal = {\apj},
         year = 2012,
        month = nov,
       volume = {760},
       number = {1},
          eid = {87},
        pages = {87},
          doi = {10.1088/0004-637X/760/1/87},
archivePrefix = {arXiv},
       eprint = {1210.3621},
 primaryClass = {astro-ph.CO},
       adsurl = {https://ui.adsabs.harvard.edu/abs/2012ApJ...760...87S}
}

@ARTICLE{Spitler2006,
       author = {{Spitler}, Lee R. and {Larsen}, S{\o}ren S. and {Strader}, Jay and {Brodie}, Jean P. and {Forbes}, Duncan A. and {Beasley}, Michael A.},
        title = "{Hubble Space Telescope ACS Wide-Field Photometry of the Sombrero Galaxy Globular Cluster System}",
      journal = {\aj},
         year = 2006,
        month = oct,
       volume = {132},
       number = {4},
        pages = {1593-1609},
          doi = {10.1086/507328},
archivePrefix = {arXiv},
       eprint = {astro-ph/0606337},
 primaryClass = {astro-ph},
       adsurl = {https://ui.adsabs.harvard.edu/abs/2006AJ....132.1593S}
}

@ARTICLE{Carretta2012,
       author = {{Carretta}, E. and {D'Orazi}, V. and {Gratton}, R.~G. and {Lucatello}, S.},
        title = "{Aluminum abundances of multiple stellar generations in the globular cluster NGC 1851}",
      journal = {\aap},
         year = 2012,
        month = jul,
       volume = {543},
          eid = {A117},
        pages = {A117},
          doi = {10.1051/0004-6361/201219277},
archivePrefix = {arXiv},
       eprint = {1205.6807},
 primaryClass = {astro-ph.GA},
       adsurl = {https://ui.adsabs.harvard.edu/abs/2012A&A...543A.117C}
}

@ARTICLE{Contursi2021,
       author = {{Contursi}, G. and {de Laverny}, P. and {Recio-Blanco}, A. and {Palicio}, P.~A.},
        title = "{GSP-spec line list for the parametrisation of Gaia-RVS stellar spectra}",
      journal = {\aap},
         year = 2021,
        month = oct,
       volume = {654},
          eid = {A130},
        pages = {A130},
          doi = {10.1051/0004-6361/202140912},
archivePrefix = {arXiv},
       eprint = {2109.06509},
 primaryClass = {astro-ph.SR},
       adsurl = {https://ui.adsabs.harvard.edu/abs/2021A&A...654A.130C}
}

@ARTICLE{Fahrion2020b,
       author = {{Fahrion}, K. and {Lyubenova}, M. and {Hilker}, M. and {van de Ven}, G. and {Falc{\'o}n-Barroso}, J. and {Leaman}, R. and {Mart{\'\i}n-Navarro}, I. and {Bittner}, A. and {Coccato}, L. and {Corsini}, E.~M. and {Gadotti}, D.~A. and {Iodice}, E. and {McDermid}, R.~M. and {Pinna}, F. and {Sarzi}, M. and {Viaene}, S. and {de Zeeuw}, P.~T. and {Zhu}, L.},
        title = "{The Fornax 3D project: Globular clusters tracing kinematics and metallicities}",
      journal = {\aap},
         year = 2020,
        month = may,
       volume = {637},
          eid = {A26},
        pages = {A26},
          doi = {10.1051/0004-6361/202037685},
archivePrefix = {arXiv},
       eprint = {2003.13705},
 primaryClass = {astro-ph.GA},
       adsurl = {https://ui.adsabs.harvard.edu/abs/2020A&A...637A..26F}
}

@ARTICLE{Diaz1989,
       author = {{Diaz}, Angeles I. and {Terlevich}, Elena and {Terlevich}, Roberto},
        title = "{Near-IR features in late type stars : their relation with stellar atmosphere parameters.}",
      journal = {\mnras},
         year = 1989,
        month = jul,
       volume = {239},
        pages = {325-345},
          doi = {10.1093/mnras/239.2.325},
       adsurl = {https://ui.adsabs.harvard.edu/abs/1989MNRAS.239..325D}
}

@ARTICLE{ArmandroffZinn1988,
       author = {{Armandroff}, Taft E. and {Zinn}, Robert},
        title = "{Integrated-Light Spectroscopy of Globular Clusters at the Infrared CA II Lines}",
      journal = {\aj},
         year = 1988,
        month = jul,
       volume = {96},
        pages = {92},
          doi = {10.1086/114792},
       adsurl = {https://ui.adsabs.harvard.edu/abs/1988AJ.....96...92A}
}

@ARTICLE{Vazdekis2010,
       author = {{Vazdekis}, A. and {S{\'a}nchez-Bl{\'a}zquez}, P. and {Falc{\'o}n-Barroso}, J. and {Cenarro}, A.~J. and {Beasley}, M.~A. and {Cardiel}, N. and {Gorgas}, J. and {Peletier}, R.~F.},
        title = "{Evolutionary stellar population synthesis with MILES - I. The base models and a new line index system}",
      journal = {\mnras},
         year = 2010,
        month = jun,
       volume = {404},
       number = {4},
        pages = {1639-1671},
          doi = {10.1111/j.1365-2966.2010.16407.x},
archivePrefix = {arXiv},
       eprint = {1004.4439},
 primaryClass = {astro-ph.CO},
       adsurl = {https://ui.adsabs.harvard.edu/abs/2010MNRAS.404.1639V}
}

@ARTICLE{Fahrion2025b,
       author = {{Fahrion}, Katja and {Beasley}, Michael A. and {Emsellem}, Eric and {Gvozdenko}, Anastasia and {M{\"u}ller}, Oliver and {Rejkuba}, Marina and {Chies-Santos}, Ana L.},
        title = "{Globular clusters in M 104: Tracing kinematics and metallicities from the centre to the halo}",
      journal = {\aap},
         year = 2025,
        month = oct,
       volume = {702},
          eid = {A59},
        pages = {A59},
          doi = {10.1051/0004-6361/202555465},
archivePrefix = {arXiv},
       eprint = {2508.10100},
 primaryClass = {astro-ph.GA},
       adsurl = {https://ui.adsabs.harvard.edu/abs/2025A&A...702A..59F}
}

@ARTICLE{Horne1986,
       author = {{Horne}, K.},
        title = "{An optimal extraction algorithm for CCD spectroscopy.}",
      journal = {\pasp},
         year = 1986,
        month = jun,
       volume = {98},
        pages = {609-617},
          doi = {10.1086/131801},
       adsurl = {https://ui.adsabs.harvard.edu/abs/1986PASP...98..609H}
}

@ARTICLE{Husser2013,
       author = {{Husser}, T.-O. and {Wende-von Berg}, S. and {Dreizler}, S. and {Homeier}, D. and {Reiners}, A. and {Barman}, T. and {Hauschildt}, P.~H.},
        title = "{A new extensive library of PHOENIX stellar atmospheres and synthetic spectra}",
      journal = {\aap},
         year = 2013,
        month = may,
       volume = {553},
          eid = {A6},
        pages = {A6},
          doi = {10.1051/0004-6361/201219058},
archivePrefix = {arXiv},
       eprint = {1303.5632},
 primaryClass = {astro-ph.SR},
       adsurl = {https://ui.adsabs.harvard.edu/abs/2013A&A...553A...6H}
}

@ARTICLE{Cappellari2023,
       author = {{Cappellari}, Michele},
        title = "{Full spectrum fitting with photometry in PPXF: stellar population versus dynamical masses, non-parametric star formation history and metallicity for 3200 LEGA-C galaxies at redshift z {\ensuremath{\approx}} 0.8}",
      journal = {\mnras},
         year = 2023,
        month = dec,
       volume = {526},
       number = {3},
        pages = {3273-3300},
          doi = {10.1093/mnras/stad2597},
archivePrefix = {arXiv},
       eprint = {2208.14974},
 primaryClass = {astro-ph.GA},
       adsurl = {https://ui.adsabs.harvard.edu/abs/2023MNRAS.526.3273C}
}

@ARTICLE{Cappellari2017,
       author = {{Cappellari}, Michele},
        title = "{Improving the full spectrum fitting method: accurate convolution with Gauss-Hermite functions}",
      journal = {\mnras},
         year = 2017,
        month = apr,
       volume = {466},
       number = {1},
        pages = {798-811},
          doi = {10.1093/mnras/stw3020},
archivePrefix = {arXiv},
       eprint = {1607.08538},
 primaryClass = {astro-ph.GA},
       adsurl = {https://ui.adsabs.harvard.edu/abs/2017MNRAS.466..798C}
}

@ARTICLE{Cappellari2004,
       author = {{Cappellari}, Michele and {Emsellem}, Eric},
        title = "{Parametric Recovery of Line-of-Sight Velocity Distributions from Absorption-Line Spectra of Galaxies via Penalized Likelihood}",
      journal = {\pasp},
         year = 2004,
        month = feb,
       volume = {116},
       number = {816},
        pages = {138-147},
          doi = {10.1086/381875},
archivePrefix = {arXiv},
       eprint = {astro-ph/0312201},
 primaryClass = {astro-ph},
       adsurl = {https://ui.adsabs.harvard.edu/abs/2004PASP..116..138C}
}

@ARTICLE{Nantais2006,
       author = {{Nantais}, Julie B. and {Huchra}, John P. and {Barmby}, Pauline and {Olsen}, Knut A.~G. and {Jarrett}, Tom H.},
        title = "{Nearby Spiral Globular Cluster Systems. I. Luminosity Functions}",
      journal = {\aj},
         year = 2006,
        month = mar,
       volume = {131},
       number = {3},
        pages = {1416-1425},
          doi = {10.1086/500011},
       adsurl = {https://ui.adsabs.harvard.edu/abs/2006AJ....131.1416N}
}

@ARTICLE{Olsen2004,
       author = {{Olsen}, Knut A.~G. and {Miller}, Bryan W. and {Suntzeff}, Nicholas B. and {Schommer}, Robert A. and {Bright}, John},
        title = "{The Globular Cluster Systems of the Sculptor Group}",
      journal = {\aj},
         year = 2004,
        month = may,
       volume = {127},
       number = {5},
        pages = {2674-2693},
          doi = {10.1086/383297},
archivePrefix = {arXiv},
       eprint = {astro-ph/0401619},
 primaryClass = {astro-ph},
       adsurl = {https://ui.adsabs.harvard.edu/abs/2004AJ....127.2674O}
}

@ARTICLE{Cantiello2018,
       author = {{Cantiello}, Michele and {Grado}, Aniello and {Rejkuba}, Marina and {Arnaboldi}, Magda and {Capaccioli}, Massimo and {Greggio}, Laura and {Iodice}, Enrica and {Limatola}, Luca},
        title = "{A VST and VISTA study of globular clusters in NGC 253}",
      journal = {\aap},
         year = 2018,
        month = mar,
       volume = {611},
          eid = {A21},
        pages = {A21},
          doi = {10.1051/0004-6361/201731325},
archivePrefix = {arXiv},
       eprint = {1711.00805},
 primaryClass = {astro-ph.GA},
       adsurl = {https://ui.adsabs.harvard.edu/abs/2018A&A...611A..21C}
}

@ARTICLE{Powalka2016,
       author = {{Powalka}, Mathieu and {Lan{\c{c}}on}, Ariane and {Puzia}, Thomas H. and {Peng}, Eric W. and {Liu}, Chengze and {Mu{\~n}oz}, Roberto P. and {Blakeslee}, John P. and {C{\^o}t{\'e}}, Patrick and {Ferrarese}, Laura and {Roediger}, Joel and {S{\'a}nchez-Janssen}, R{\'u}ben and {Zhang}, Hongxin and {Durrell}, Patrick R. and {Cuillandre}, Jean-Charles and {Duc}, Pierre-Alain and {Guhathakurta}, Puragra and {Gwyn}, S.~D.~J. and {Hudelot}, Patrick and {Mei}, Simona and {Toloba}, Elisa},
        title = "{The Next Generation Virgo Cluster Survey (NGVS). XXV. Fiducial Panchromatic Colors of Virgo Core Globular Clusters and Their Comparison to Model Predictions}",
      journal = {\apjs},
         year = 2016,
        month = nov,
       volume = {227},
       number = {1},
          eid = {12},
        pages = {12},
          doi = {10.3847/0067-0049/227/1/12},
archivePrefix = {arXiv},
       eprint = {1608.04742},
 primaryClass = {astro-ph.GA},
       adsurl = {https://ui.adsabs.harvard.edu/abs/2016ApJS..227...12P}
}

@ARTICLE{Peacock2015,
       author = {{Peacock}, Mark B. and {Strader}, Jay and {Romanowsky}, Aaron J. and {Brodie}, Jean P.},
        title = "{Detection of a Distinct Metal-poor Stellar Halo in the Early-type Galaxy NGC 3115{\textdagger}}",
      journal = {\apj},
         year = 2015,
        month = feb,
       volume = {800},
       number = {1},
          eid = {13},
        pages = {13},
          doi = {10.1088/0004-637X/800/1/13},
archivePrefix = {arXiv},
       eprint = {1412.2752},
 primaryClass = {astro-ph.GA},
       adsurl = {https://ui.adsabs.harvard.edu/abs/2015ApJ...800...13P}
}

@ARTICLE{Jensen2025,
       author = {{Jensen}, Joseph B. and {Blakeslee}, John P. and {Cantiello}, Michele and {Cowles}, Mikaela and {Anand}, Gagandeep S. and {Tully}, R. Brent and {Kourkchi}, Ehsan and {Raimondo}, Gabriella},
        title = "{The TRGB{\ensuremath{-}}SBF Project. III. Refining the HST Surface Brightness Fluctuation Distance Scale Calibration with JWST}",
      journal = {\apj},
         year = 2025,
        month = jul,
       volume = {987},
       number = {1},
          eid = {87},
        pages = {87},
          doi = {10.3847/1538-4357/addfd6},
archivePrefix = {arXiv},
       eprint = {2502.15935},
 primaryClass = {astro-ph.CO},
       adsurl = {https://ui.adsabs.harvard.edu/abs/2025ApJ...987...87J}
}

@ARTICLE{Anand2025_Virgo,
       author = {{Anand}, Gagandeep S. and {Tully}, R. Brent and {Cohen}, Yotam and {Shaya}, Edward J. and {Makarov}, Dmitry I. and {Makarova}, Lidia N. and {Chazov}, Maksim I. and {Blakeslee}, John P. and {Cantiello}, Michele and {Jensen}, Joseph B. and {Kourkchi}, Ehsan and {Raimondo}, Gabriella},
        title = "{The TRGB─SBF Project. II. Resolving the Virgo Cluster with JWST}",
      journal = {\apj},
         year = 2025,
        month = mar,
       volume = {982},
       number = {1},
          eid = {26},
        pages = {26},
          doi = {10.3847/1538-4357/adb399},
archivePrefix = {arXiv},
       eprint = {2408.16810},
 primaryClass = {astro-ph.GA},
       adsurl = {https://ui.adsabs.harvard.edu/abs/2025ApJ...982...26A}
}

@ARTICLE{Anand2024_Fornax,
       author = {{Anand}, Gagandeep S. and {Tully}, R. Brent and {Cohen}, Yotam and {Makarov}, Dmitry I. and {Makarova}, Lidia N. and {Jensen}, Joseph B. and {Blakeslee}, John P. and {Cantiello}, Michele and {Kourkchi}, Ehsan and {Raimondo}, Gabriella},
        title = "{The TRGB-SBF Project. I. A Tip of the Red Giant Branch Distance to the Fornax Cluster with JWST}",
      journal = {\apj},
         year = 2024,
        month = oct,
       volume = {973},
       number = {2},
          eid = {83},
        pages = {83},
          doi = {10.3847/1538-4357/ad64c7},
archivePrefix = {arXiv},
       eprint = {2405.03743},
 primaryClass = {astro-ph.CO},
       adsurl = {https://ui.adsabs.harvard.edu/abs/2024ApJ...973...83A}
}

@ARTICLE{Bailin2010,
       author = {{Bailin}, Jeremy and {Bell}, Eric F. and {Chappell}, Samantha N. and {Radburn-Smith}, David J. and {de Jong}, Roelof S.},
        title = "{The Resolved Stellar Halo of NGC 253}",
      journal = {\apj},
         year = 2011,
        month = jul,
       volume = {736},
       number = {1},
          eid = {24},
        pages = {24},
          doi = {10.1088/0004-637X/736/1/24},
archivePrefix = {arXiv},
       eprint = {1105.0005},
 primaryClass = {astro-ph.CO},
       adsurl = {https://ui.adsabs.harvard.edu/abs/2011ApJ...736...24B}
}

@ARTICLE{Guerou2016,
       author = {{Gu{\'e}rou}, A. and {Emsellem}, E. and {Krajnovi{\'c}}, D. and {McDermid}, R.~M. and {Contini}, T. and {Weilbacher}, P.~M.},
        title = "{Exploring the mass assembly of the early-type disc galaxy NGC 3115 with MUSE}",
      journal = {\aap},
         year = 2016,
        month = jul,
       volume = {591},
          eid = {A143},
        pages = {A143},
          doi = {10.1051/0004-6361/201628743},
archivePrefix = {arXiv},
       eprint = {1605.07667},
 primaryClass = {astro-ph.GA},
       adsurl = {https://ui.adsabs.harvard.edu/abs/2016A&A...591A.143G}
}

@ARTICLE{Cappellari2013,
       author = {{Cappellari}, Michele and {Scott}, Nicholas and {Alatalo}, Katherine and {Blitz}, Leo and {Bois}, Maxime and {Bournaud}, Fr{\'e}d{\'e}ric and {Bureau}, M. and {Crocker}, Alison F. and {Davies}, Roger L. and {Davis}, Timothy A. and {de Zeeuw}, P.~T. and {Duc}, Pierre-Alain and {Emsellem}, Eric and {Khochfar}, Sadegh and {Krajnovi{\'c}}, Davor and {Kuntschner}, Harald and {McDermid}, Richard M. and {Morganti}, Raffaella and {Naab}, Thorsten and {Oosterloo}, Tom and {Sarzi}, Marc and {Serra}, Paolo and {Weijmans}, Anne-Marie and {Young}, Lisa M.},
        title = "{The ATLAS$^{3D}$ project - XV. Benchmark for early-type galaxies scaling relations from 260 dynamical models: mass-to-light ratio, dark matter, Fundamental Plane and Mass Plane}",
      journal = {\mnras},
         year = 2013,
        month = jul,
       volume = {432},
       number = {3},
        pages = {1709-1741},
          doi = {10.1093/mnras/stt562},
archivePrefix = {arXiv},
       eprint = {1208.3522},
 primaryClass = {astro-ph.CO},
       adsurl = {https://ui.adsabs.harvard.edu/abs/2013MNRAS.432.1709C}
}

@ARTICLE{OSullivan2018,
       author = {{O'Sullivan}, E. and {Combes}, F. and {Salom{\'e}}, P. and {David}, L.~P. and {Babul}, A. and {Vrtilek}, J.~M. and {Lim}, J. and {Olivares}, V. and {Raychaudhury}, S. and {Schellenberger}, G.},
        title = "{Cold gas in a complete sample of group-dominant early-type galaxies}",
      journal = {\aap},
         year = 2018,
        month = oct,
       volume = {618},
          eid = {A126},
        pages = {A126},
          doi = {10.1051/0004-6361/201833580},
archivePrefix = {arXiv},
       eprint = {1807.09110},
 primaryClass = {astro-ph.GA},
       adsurl = {https://ui.adsabs.harvard.edu/abs/2018A&A...618A.126O}
}

@ARTICLE{Liu2019,
       author = {{Liu}, Yiqing and {Peng}, Eric W. and {Jord{\'a}n}, Andr{\'e}s and {Blakeslee}, John P. and {C{\^o}t{\'e}}, Patrick and {Ferrarese}, Laura and {Puzia}, Thomas H.},
        title = "{The ACS Fornax Cluster Survey. III. Globular Cluster Specific Frequencies of Early-type Galaxies}",
      journal = {\apj},
         year = 2019,
        month = apr,
       volume = {875},
       number = {2},
          eid = {156},
        pages = {156},
          doi = {10.3847/1538-4357/ab12d9},
archivePrefix = {arXiv},
       eprint = {1904.06909},
 primaryClass = {astro-ph.GA},
       adsurl = {https://ui.adsabs.harvard.edu/abs/2019ApJ...875..156L}
}

@ARTICLE{Peng2008,
       author = {{Peng}, Eric W. and {Jord{\'a}n}, Andr{\'e}s and {C{\^o}t{\'e}}, Patrick and {Takamiya}, Marianne and {West}, Michael J. and {Blakeslee}, John P. and {Chen}, Chin-Wei and {Ferrarese}, Laura and {Mei}, Simona and {Tonry}, John L. and {West}, Andrew A.},
        title = "{The ACS Virgo Cluster Survey. XV. The Formation Efficiencies of Globular Clusters in Early-Type Galaxies: The Effects of Mass and Environment}",
      journal = {\apj},
         year = 2008,
        month = jul,
       volume = {681},
       number = {1},
        pages = {197-224},
          doi = {10.1086/587951},
archivePrefix = {arXiv},
       eprint = {0803.0330},
 primaryClass = {astro-ph},
       adsurl = {https://ui.adsabs.harvard.edu/abs/2008ApJ...681..197P}
}

@ARTICLE{Paturel1992,
       author = {{Paturel}, G. and {Garnier}, R.},
        title = "{The use of velocity dispersion in globular clusters as a distance indicator.}",
      journal = {\aap},
         year = 1992,
        month = feb,
       volume = {254},
        pages = {93-95},
       adsurl = {https://ui.adsabs.harvard.edu/abs/1992A&A...254...93P}
}

@ARTICLE{Strader2009,
       author = {{Strader}, Jay and {Smith}, Graeme H. and {Larsen}, Soeren and {Brodie}, Jean P. and {Huchra}, John P.},
        title = "{Mass-to-Light Ratios for M31 Globular Clusters: Age Dating and a Surprising Metallicity Trend}",
      journal = {\aj},
         year = 2009,
        month = aug,
       volume = {138},
       number = {2},
        pages = {547-557},
          doi = {10.1088/0004-6256/138/2/547},
archivePrefix = {arXiv},
       eprint = {0906.0397},
 primaryClass = {astro-ph.GA},
       adsurl = {https://ui.adsabs.harvard.edu/abs/2009AJ....138..547S}
}

@ARTICLE{Djorgovski1995,
       author = {{Djorgovski}, S.},
        title = "{The Fundamental Plane Correlations for Globular Clusters}",
      journal = {\apjl},
         year = 1995,
        month = jan,
       volume = {438},
        pages = {L29},
          doi = {10.1086/187707},
       adsurl = {https://ui.adsabs.harvard.edu/abs/1995ApJ...438L..29D}
}

@ARTICLE{Gratton2019,
       author = {{Gratton}, Raffaele and {Bragaglia}, Angela and {Carretta}, Eugenio and {D'Orazi}, Valentina and {Lucatello}, Sara and {Sollima}, Antonio},
        title = "{What is a globular cluster? An observational perspective}",
      journal = {\aapr},
         year = 2019,
        month = nov,
       volume = {27},
       number = {1},
          eid = {8},
        pages = {8},
          doi = {10.1007/s00159-019-0119-3},
archivePrefix = {arXiv},
       eprint = {1911.02835},
 primaryClass = {astro-ph.SR},
       adsurl = {https://ui.adsabs.harvard.edu/abs/2019A&ARv..27....8G}
}

@ARTICLE{Clontz2024,
       author = {{Clontz}, C. and {Seth}, A.~C. and {Dotter}, A. and {H{\"a}berle}, M. and {Nitschai}, M.~S. and {Neumayer}, N. and {Feldmeier-Krause}, A. and {Latour}, M. and {Wang}, Z. and {Souza}, S.~O. and {Kacharov}, N. and {Bellini}, A. and {Libralato}, M. and {Pechetti}, R. and {van de Ven}, G. and {Alfaro-Cuello}, M.},
        title = "{oMEGACat. IV. Constraining the Ages of Omega Centauri Subgiant Branch Stars with HST and MUSE}",
      journal = {\apj},
         year = 2024,
        month = dec,
       volume = {977},
       number = {1},
          eid = {14},
        pages = {14},
          doi = {10.3847/1538-4357/ad8621},
archivePrefix = {arXiv},
       eprint = {2409.13855},
 primaryClass = {astro-ph.GA},
       adsurl = {https://ui.adsabs.harvard.edu/abs/2024ApJ...977...14C}
}

@ARTICLE{Beasley2018,
       author = {{Beasley}, Michael A. and {Trujillo}, Ignacio and {Leaman}, Ryan and {Montes}, Mireia},
        title = "{A single population of red globular clusters around the massive compact galaxy NGC 1277}",
      journal = {\nat},
         year = 2018,
        month = mar,
       volume = {555},
       number = {7697},
        pages = {483-486},
          doi = {10.1038/nature25756},
archivePrefix = {arXiv},
       eprint = {1803.04893},
 primaryClass = {astro-ph.GA},
       adsurl = {https://ui.adsabs.harvard.edu/abs/2018Natur.555..483B}
}

@ARTICLE{FerreMateu2021,
       author = {{Ferr{\'e}-Mateu}, A. and {Mezcua}, M. and {Barrows}, R.~S.},
        title = "{A search for active galactic nuclei in low-mass compact galaxies}",
      journal = {\mnras},
         year = 2021,
        month = oct,
       volume = {506},
       number = {4},
        pages = {4702-4714},
          doi = {10.1093/mnras/stab1915},
archivePrefix = {arXiv},
       eprint = {2107.02141},
 primaryClass = {astro-ph.GA},
       adsurl = {https://ui.adsabs.harvard.edu/abs/2021MNRAS.506.4702F}
}

@ARTICLE{BurkertForbes2020,
       author = {{Burkert}, Andreas and {Forbes}, Duncan A.},
        title = "{High-precision Dark Halo Virial Masses from Globular Cluster Numbers: Implications for Globular Cluster Formation and Galaxy Assembly}",
      journal = {\aj},
         year = 2020,
        month = feb,
       volume = {159},
       number = {2},
          eid = {56},
        pages = {56},
          doi = {10.3847/1538-3881/ab5b0e},
archivePrefix = {arXiv},
       eprint = {1901.00900},
 primaryClass = {astro-ph.GA},
       adsurl = {https://ui.adsabs.harvard.edu/abs/2020AJ....159...56B}
}

@ARTICLE{Forbes2022,
       author = {{Forbes}, Duncan A. and {Ferr{\'e}-Mateu}, Anna and {Gannon}, Jonah S. and {Romanowsky}, Aaron J. and {Carlin}, Jeffrey L. and {Brodie}, Jean P. and {Day}, Jacob},
        title = "{Low-metallicity globular clusters in the low-mass isolated spiral galaxy NGC 2403}",
      journal = {\mnras},
         year = 2022,
        month = may,
       volume = {512},
       number = {1},
        pages = {802-810},
          doi = {10.1093/mnras/stac503},
archivePrefix = {arXiv},
       eprint = {2202.10539},
 primaryClass = {astro-ph.GA},
       adsurl = {https://ui.adsabs.harvard.edu/abs/2022MNRAS.512..802F}
}

@ARTICLE{Revnivtsev2008,
       author = {{Revnivtsev}, M. and {Churazov}, E. and {Sazonov}, S. and {Forman}, W. and {Jones}, C.},
        title = "{Universal X-ray emissivity of the stellar population in early-type galaxies: unresolved X-ray sources in NGC 3379}",
      journal = {\aap},
         year = 2008,
        month = oct,
       volume = {490},
       number = {1},
        pages = {37-43},
          doi = {10.1051/0004-6361:200809889},
archivePrefix = {arXiv},
       eprint = {0804.0319},
 primaryClass = {astro-ph},
       adsurl = {https://ui.adsabs.harvard.edu/abs/2008A&A...490...37R}
}

@ARTICLE{Dooley2017,
       author = {{Dooley}, Gregory A. and {Peter}, Annika H.~G. and {Carlin}, Jeffrey L. and {Frebel}, Anna and {Bechtol}, Keith and {Willman}, Beth},
        title = "{The predicted luminous satellite populations around SMC- and LMC-mass galaxies - a missing satellite problem around the LMC?}",
      journal = {\mnras},
         year = 2017,
        month = nov,
       volume = {472},
       number = {1},
        pages = {1060-1073},
          doi = {10.1093/mnras/stx2001},
archivePrefix = {arXiv},
       eprint = {1703.05321},
 primaryClass = {astro-ph.GA},
       adsurl = {https://ui.adsabs.harvard.edu/abs/2017MNRAS.472.1060D}
}

@ARTICLE{Karachentsev2020,
       author = {{Karachentsev}, Igor D. and {Makarova}, Lidia N. and {Brent Tully}, R. and {Anand}, Gagandeep S. and {Rizzi}, Luca and {Shaya}, Edward J.},
        title = "{Distance and mass of the M 104 (Sombrero) group}",
      journal = {\aap},
         year = 2020,
        month = nov,
       volume = {643},
          eid = {A124},
        pages = {A124},
          doi = {10.1051/0004-6361/202038928},
archivePrefix = {arXiv},
       eprint = {2008.13152},
 primaryClass = {astro-ph.GA},
       adsurl = {https://ui.adsabs.harvard.edu/abs/2020A&A...643A.124K}
}

@ARTICLE{Jensen2003,
       author = {{Jensen}, Joseph B. and {Tonry}, John L. and {Barris}, Brian J. and {Thompson}, Rodger I. and {Liu}, Michael C. and {Rieke}, Marcia J. and {Ajhar}, Edward A. and {Blakeslee}, John P.},
        title = "{Measuring Distances and Probing the Unresolved Stellar Populations of Galaxies Using Infrared Surface Brightness Fluctuations}",
      journal = {\apj},
         year = 2003,
        month = feb,
       volume = {583},
       number = {2},
        pages = {712-726},
          doi = {10.1086/345430},
archivePrefix = {arXiv},
       eprint = {astro-ph/0210129},
 primaryClass = {astro-ph},
       adsurl = {https://ui.adsabs.harvard.edu/abs/2003ApJ...583..712J}
}

@ARTICLE{Usher2018,
       author = {{Usher}, Christopher and {Pfeffer}, Joel and {Bastian}, Nate and {Kruijssen}, J.~M. Diederik and {Crain}, Robert A. and {Reina-Campos}, Marta},
        title = "{The origin of the `blue tilt' of globular cluster populations in the E-MOSAICS simulations}",
      journal = {\mnras},
         year = 2018,
        month = nov,
       volume = {480},
       number = {3},
        pages = {3279-3301},
          doi = {10.1093/mnras/sty1895},
archivePrefix = {arXiv},
       eprint = {1807.03084},
 primaryClass = {astro-ph.GA},
       adsurl = {https://ui.adsabs.harvard.edu/abs/2018MNRAS.480.3279U}
}

@ARTICLE{Harris2006,
       author = {{Harris}, William E. and {Harris}, Gretchen L.~H. and {Barmby}, Pauline and {McLaughlin}, Dean E. and {Forbes}, Duncan A.},
        title = "{Structural Parameters for Globular Clusters in NGC 5128. II. Hubble Space Telescope ACS Imaging and New Clusters}",
      journal = {\aj},
         year = 2006,
        month = nov,
       volume = {132},
       number = {5},
        pages = {2187-2197},
          doi = {10.1086/507579},
archivePrefix = {arXiv},
       eprint = {astro-ph/0607269},
 primaryClass = {astro-ph},
       adsurl = {https://ui.adsabs.harvard.edu/abs/2006AJ....132.2187H}
}

@ARTICLE{Fahrion2026a,
       author = {{Fahrion}, Katja and {Beasley}, Michael A. and {Gvozdenko}, Anastasia and {van de Ven}, Glenn and {Rhode}, Katherine L. and {Chies-Santos}, Ana L. and {Ferre-Mateu}, Anna and {Rejkuba}, Marina and {M{\"u}ller}, Oliver and {Emsellem}, Eric},
        title = "{Distance measurements from the internal dynamics of globular clusters: Application to the Sombrero galaxy (M 104)}",
      journal = {The Open Journal of Astrophysics},
         year = 2026,
        month = jan,
       volume = {9},
        pages = {55146},
          doi = {10.33232/001c.155146},
archivePrefix = {arXiv},
       eprint = {2509.14948},
 primaryClass = {astro-ph.GA},
       adsurl = {https://ui.adsabs.harvard.edu/abs/2026OJAp....955146F}
}

@ARTICLE{Beasley2025,
       author = {{Beasley}, M.~A. and {Fahrion}, K. and {Guerra Arencibia}, S. and {Gvozdenko}, A. and {Montes}, M.},
        title = "{A new way to measure the distance to NGC1052-DF2}",
      journal = {\aap},
         year = 2025,
        month = may,
       volume = {697},
          eid = {A144},
        pages = {A144},
          doi = {10.1051/0004-6361/202452446},
archivePrefix = {arXiv},
       eprint = {2503.03403},
 primaryClass = {astro-ph.GA},
       adsurl = {https://ui.adsabs.harvard.edu/abs/2025A&A...697A.144B}
}

@ARTICLE{Beasley2024,
       author = {{Beasley}, Michael A. and {Fahrion}, Katja and {Gvozdenko}, Anastasia},
        title = "{Measuring distances to galaxies with globular cluster velocity dispersions}",
      journal = {\mnras},
         year = 2024,
        month = jan,
       volume = {527},
       number = {3},
        pages = {5767-5775},
          doi = {10.1093/mnras/stad3541},
archivePrefix = {arXiv},
       eprint = {2312.01420},
 primaryClass = {astro-ph.GA},
       adsurl = {https://ui.adsabs.harvard.edu/abs/2024MNRAS.527.5767B}
}

@ARTICLE{Harris2010,
       author = {{Harris}, William E. and {Spitler}, Lee R. and {Forbes}, Duncan A. and {Bailin}, Jeremy},
        title = "{Diamonds on the Hat: globular clusters in the Sombrero galaxy (M104)}",
      journal = {\mnras},
         year = 2010,
        month = jan,
       volume = {401},
       number = {3},
        pages = {1965-1982},
          doi = {10.1111/j.1365-2966.2009.15783.x},
archivePrefix = {arXiv},
       eprint = {0909.4805},
 primaryClass = {astro-ph.GA},
       adsurl = {https://ui.adsabs.harvard.edu/abs/2010MNRAS.401.1965H}
}

@ARTICLE{Peng2006,
       author = {{Peng}, Eric W. and {Jord{\'a}n}, Andr{\'e}s and {C{\^o}t{\'e}}, Patrick and {Blakeslee}, John P. and {Ferrarese}, Laura and {Mei}, Simona and {West}, Michael J. and {Merritt}, David and {Milosavljevi{\'c}}, Milos and {Tonry}, John L.},
        title = "{The ACS Virgo Cluster Survey. IX. The Color Distributions of Globular Cluster Systems in Early-Type Galaxies}",
      journal = {\apj},
         year = 2006,
        month = mar,
       volume = {639},
       number = {1},
        pages = {95-119},
          doi = {10.1086/498210},
archivePrefix = {arXiv},
       eprint = {astro-ph/0509654},
 primaryClass = {astro-ph},
       adsurl = {https://ui.adsabs.harvard.edu/abs/2006ApJ...639...95P}
}

@ARTICLE{Veljanoski2013,
       author = {{Veljanoski}, J. and {Ferguson}, A.~M.~N. and {Mackey}, A.~D. and {Huxor}, A.~P. and {Irwin}, M.~J. and {C{\^o}t{\'e}}, P. and {Tanvir}, N.~R. and {Bernard}, E.~J. and {Chapman}, S.~C. and {Ibata}, R.~A. and {Fardal}, M. and {Lewis}, G.~F. and {Martin}, N.~F. and {McConnachie}, A. and {Pe{\~n}arrubia}, J.},
        title = "{Kinematics of Outer Halo Globular Clusters in M31}",
      journal = {\apjl},
         year = 2013,
        month = may,
       volume = {768},
       number = {2},
          eid = {L33},
        pages = {L33},
          doi = {10.1088/2041-8205/768/2/L33},
archivePrefix = {arXiv},
       eprint = {1303.7368},
 primaryClass = {astro-ph.GA},
       adsurl = {https://ui.adsabs.harvard.edu/abs/2013ApJ...768L..33V}
}

@ARTICLE{Martini2004,
       author = {{Martini}, Paul and {Ho}, Luis C.},
        title = "{A Population of Massive Globular Clusters in NGC 5128}",
      journal = {\apj},
         year = 2004,
        month = jul,
       volume = {610},
       number = {1},
        pages = {233-246},
          doi = {10.1086/421458},
archivePrefix = {arXiv},
       eprint = {astro-ph/0404003},
 primaryClass = {astro-ph},
       adsurl = {https://ui.adsabs.harvard.edu/abs/2004ApJ...610..233M}
}

@ARTICLE{Hernandez2018,
       author = {{Hernandez}, Svea and {Larsen}, S{\o}ren and {Trager}, Scott and {Kaper}, Lex and {Groot}, Paul},
        title = "{Chemical abundances of globular clusters in NGC 5128 (Centaurus A)}",
      journal = {\mnras},
         year = 2018,
        month = jun,
       volume = {476},
       number = {4},
        pages = {5189-5215},
          doi = {10.1093/mnras/sty519},
archivePrefix = {arXiv},
       eprint = {1802.09534},
 primaryClass = {astro-ph.GA},
       adsurl = {https://ui.adsabs.harvard.edu/abs/2018MNRAS.476.5189H}
}

@ARTICLE{Fellhauer2002,
       author = {{Fellhauer}, Michael and {Kroupa}, Pavel},
        title = "{The formation of ultracompact dwarf galaxies}",
      journal = {\mnras},
         year = 2002,
        month = mar,
       volume = {330},
       number = {3},
        pages = {642-650},
          doi = {10.1046/j.1365-8711.2002.05087.x},
archivePrefix = {arXiv},
       eprint = {astro-ph/0110621},
 primaryClass = {astro-ph},
       adsurl = {https://ui.adsabs.harvard.edu/abs/2002MNRAS.330..642F}
}

@ARTICLE{Jordan2009,
       author = {{Jord{\'a}n}, Andr{\'e}s and {Peng}, Eric W. and {Blakeslee}, John P. and {C{\^o}t{\'e}}, Patrick and {Eyheramendy}, Susana and {Ferrarese}, Laura and {Mei}, Simona and {Tonry}, John L. and {West}, Michael J.},
        title = "{The ACS Virgo Cluster Survey XVI. Selection Procedure and Catalogs of Globular Cluster Candidates}",
      journal = {\apjs},
         year = 2009,
        month = jan,
       volume = {180},
       number = {1},
        pages = {54-66},
          doi = {10.1088/0067-0049/180/1/54},
       adsurl = {https://ui.adsabs.harvard.edu/abs/2009ApJS..180...54J}
}

@ARTICLE{Dabringhausen2010,
       author = {{Dabringhausen}, J. and {Fellhauer}, M. and {Kroupa}, P.},
        title = "{Mass loss and expansion of ultra compact dwarf galaxies through gas expulsion and stellar evolution for top-heavy stellar initial mass functions}",
      journal = {\mnras},
         year = 2010,
        month = apr,
       volume = {403},
       number = {2},
        pages = {1054-1071},
          doi = {10.1111/j.1365-2966.2009.16189.x},
archivePrefix = {arXiv},
       eprint = {0912.2998},
 primaryClass = {astro-ph.CO},
       adsurl = {https://ui.adsabs.harvard.edu/abs/2010MNRAS.403.1054D}
}

@ARTICLE{Tremou2018,
       author = {{Tremou}, Evangelia and {Strader}, Jay and {Chomiuk}, Laura and {Shishkovsky}, Laura and {Maccarone}, Thomas J. and {Miller-Jones}, James C.~A. and {Tudor}, Vlad and {Heinke}, Craig O. and {Sivakoff}, Gregory R. and {Seth}, Anil C. and {Noyola}, Eva},
        title = "{The MAVERIC Survey: Still No Evidence for Accreting Intermediate-mass Black Holes in Globular Clusters}",
      journal = {\apj},
         year = 2018,
        month = jul,
       volume = {862},
       number = {1},
          eid = {16},
        pages = {16},
          doi = {10.3847/1538-4357/aac9b9},
archivePrefix = {arXiv},
       eprint = {1806.00259},
 primaryClass = {astro-ph.HE},
       adsurl = {https://ui.adsabs.harvard.edu/abs/2018ApJ...862...16T}
}

@ARTICLE{Baumgardt2019,
       author = {{Baumgardt}, H. and {He}, C. and {Sweet}, S.~M. and {Drinkwater}, M. and {Sollima}, A. and {Hurley}, J. and {Usher}, C. and {Kamann}, S. and {Dalgleish}, H. and {Dreizler}, S. and {Husser}, T.-O.},
        title = "{No evidence for intermediate-mass black holes in the globular clusters {\ensuremath{\omega}} Cen and NGC 6624}",
      journal = {\mnras},
         year = 2019,
        month = oct,
       volume = {488},
       number = {4},
        pages = {5340-5351},
          doi = {10.1093/mnras/stz2060},
archivePrefix = {arXiv},
       eprint = {1907.10845},
 primaryClass = {astro-ph.GA},
       adsurl = {https://ui.adsabs.harvard.edu/abs/2019MNRAS.488.5340B}
}

@ARTICLE{Forbes2014,
       author = {{Forbes}, Duncan A. and {Norris}, Mark A. and {Strader}, Jay and {Romanowsky}, Aaron J. and {Pota}, Vincenzo and {Kannappan}, Sheila J. and {Brodie}, Jean P. and {Huxor}, Avon},
        title = "{The AIMSS Project II: dynamical-to-stellar mass ratios across the star cluster-galaxy divide}",
      journal = {\mnras},
         year = 2014,
        month = nov,
       volume = {444},
       number = {4},
        pages = {2993-3003},
          doi = {10.1093/mnras/stu1631},
archivePrefix = {arXiv},
       eprint = {1408.2558},
 primaryClass = {astro-ph.GA},
       adsurl = {https://ui.adsabs.harvard.edu/abs/2014MNRAS.444.2993F}
}

@ARTICLE{Mieske2013,
       author = {{Mieske}, S. and {Frank}, M.~J. and {Baumgardt}, H. and {L{\"u}tzgendorf}, N. and {Neumayer}, N. and {Hilker}, M.},
        title = "{On central black holes in ultra-compact dwarf galaxies}",
      journal = {\aap},
         year = 2013,
        month = oct,
       volume = {558},
          eid = {A14},
        pages = {A14},
          doi = {10.1051/0004-6361/201322167},
archivePrefix = {arXiv},
       eprint = {1308.1398},
 primaryClass = {astro-ph.CO},
       adsurl = {https://ui.adsabs.harvard.edu/abs/2013A&A...558A..14M}
}

\begin{appendix}
\nolinenumbers 

    \section{Input globular cluster catalogues}
    \label{app:literature_catalogues}
    Table \ref{tab:catalogues} lists the photometric and spectroscopic catalogues that we used to prepare the observations. We note that additional catalogues might be available in the literature to be used in future work.

    \subsection{Matching with literature catalogues}
    To prepare the FLAMES observations and collect photometric and spectroscopic properties of our GC sample, we downloaded the catalogues listed in Table \ref{tab:catalogues}. To match with the sky coordinates of our observations we used a 1\arcsec\,matching radius. 
    In addition to the listed catalogues, we also used Gaia DR3 catalogues \citep{GaiaCollaboration2023} of sources around each target galaxy to select additional GCs and collect Gaia photometry.

    Due to the low number of catalogued GCs available in the Sculptor galaxies, we further used the publicly available VST-ATLAS catalogues \citep{Shanks2013} to select GC candidates based on colours. These provide OmegaCam photometry in the $ugriz$ system for bright sources.
    
    \subsection{NGC\,4697 as a special case}
    We used available HST-based GC catalogues for most galaxies in the Virgo cluster from the Advanced Camera for Surveys Virgo Cluster Survey (ACSVCS) project \citep{Jordan2009}. NGC\,4697 did not have such a catalogue available, but this galaxy was covered with HST ACS in the same F475W and F850LP filters that ACSVCS used in multiple pointings (HST proposal ID 10003, PI: C. Sarazin). To obtain GC candidates in this galaxy, we used a machine learning model that was originally trained on ACSVCS and ACS Fornax Cluster Survey GC data (ACSFCS; \citealt{Jordan2015}) as described in \cite{Globsi}. 
    
    To apply the model, we downloaded the science-ready HST data from the ESA Hubble Science Archive\footnote{\url{https://hst.esac.esa.int}}, and performed source finding using \textsc{photutils} \citep{photutils}. As in \cite{Globsi}, we derived basic photometric properties of all detected sources including aperture magnitudes and colours, which were then fed into the trained random forest model. The output was a list of high probability GCs that we used to prepare our observations with FLAMES.

    Of 331 identified GC candidates with $g < 24$ mag, 82 were observed. Of those 72 had sufficient signal-to-noise to measure velocities, and 65 (90\%) of them have velocities $v > 400$ km s$^{-1}$, identifying them as GCs of NGC\,4697.

\begin{table*}[b]
    \caption{Overview of photometric and spectroscopic GC catalogues}
    \centering
    \begin{tabular}{c| l l | l }\hline\hline
    Galaxy & Photometric catalogues & Comments & Spectroscopic catalogues \\ \hline 
    M104 & \cite{Spitler2006, Harris2010} & HST/ACS F475W, F814W & \cite{Dowell2014} \\
            & \cite{RhodeZepf2004} & BVR & \cite{AlvesBrito2011} \\
            & \cite{Kang2020} & ugi & \\ \hline
    M105 & \cite{Whitlock2003} & UBRI & \cite{Bergond2006}\\
            & \cite{Faifer2011} & gri \\
            & \cite{RhodeZepf2004} & BVR \\ \hline
    M49 & \cite{Jordan2009} & HST/ACS F475W, F850LP & \cite{Cote2003} \\
        & \cite{Rhode2001} & BVR & \cite{Sharples1998} \\
         & \cite{Liu2015} & UCDs, ugriz Ks & \\ \hline
    M60 & \cite{Jordan2009} & HST/ACS F475W, F850LP & \cite{Pota2015} \\
        & \cite{Strader2012} & HST/ACS F475W, F850LP & \cite{Forbes2017} \\
        & \cite{Liu2015, Liu2020} & UCDs, ugriz Ks & \\ \hline
    M84  & \cite{Jordan2009} & HST/ACS F475W, F850LP & \cite{Forbes2017} \\
    M86        & \cite{RhodeZepf2004} & BVR & \\
                & \cite{Powalka2016} & ugriz Ks & \\ \hline
    M87 & \cite{Jordan2009} & HST/ACS F475W, F850LP & \cite{Forbes2017} \\
         & \cite{Powalka2016} & ugriz Ks & \\
         & \cite{Oldham2015} & ugriz & \\
         & \cite{Liu2015, Liu2020} & UCDs, ugriz Ks & \\ \hline
    M89 & \cite{Jordan2009} & HST/ACS F475W, F850LP & \cite{Forbes2017} \\
        & \cite{ChiesSantos2011} & & \\
        & \cite{Liu2015, Liu2020} & UCDs, ugriz Ks & \\ \hline
    NGC\,4636 & \cite{Dirsch2005} & C T1 & \cite{Schuberth2006} \\
            &  & & \cite{Park2010} \\ \hline
    NGC\,4697 & GlobsML, \cite{Globsi} & HST/ACS F475W, F850LP & \cite{Forbes2017} \\ \hline
    NGC\,1399 & \cite{Jordan2015} & HST/ACS F475W, F850LP &  \cite{Chaturvedi2022} \\
    NGC\,1404 & \cite{LomeliNunez2025} & ugriz & \\
    NGC\,1380 & \cite{Cantiello2020} & ugriz & \\ \hline
    NGC\,253 & \cite{Cantiello2018} & ugriz & \cite{Olsen2004} \\
             & \cite{Nantais2006} & JHK & \\ \hline
    NGC\,300 & \cite{Fielder2025} & & \cite{Fielder2025} \\
           & \cite{Nantais2006} & JHK & \cite{Olsen2004} \\
    NGC\,7793 & VST-ATLAS, \cite{Shanks2013} & ugriz & \\
    NGC\,55 &  &  & \\ \hline
    \end{tabular}
    \label{tab:catalogues}
\end{table*}

\section{Example pointing setup}
\label{app:pointing_setup}
\begin{figure}
    \centering
    \includegraphics[width=0.45\textwidth]{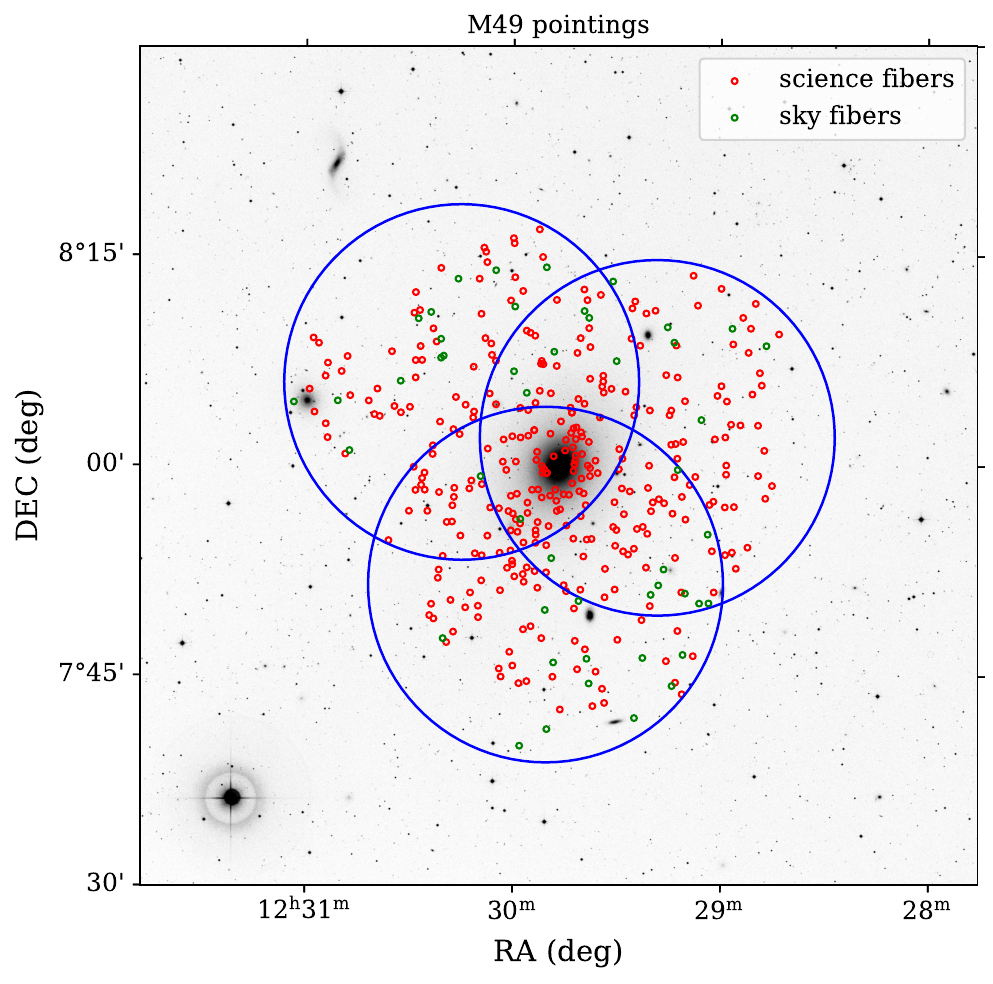}
    \caption{Example of observation setup shown for M49. The blue circles show the FLAMES field of view (25\arcmin\,diameter) for the three fields overplotted on a DSS2 image of M49 in greyscale. Red and green circles show the positions of science and sky fibres. The sizes of the small circles are not to scale.}
    \label{fig:M49_pointings}
\end{figure}

We show the pointing setup as an example for M49 in Figure \ref{fig:M49_pointings}. The figure shows the sky position of the three FLAMES fields (25\arcmin\,diameter) as well as the positions of science and sky fibres. 

\section{Velocity catalogue}
\label{app:vel_catalogue}
We make our catalogue of LOS velocities publicly available for the 1062 GCs and 101 UCDs in our initial selection (see Sect. \ref{sect:selecting}), and report the velocities of foreground stars. Table \ref{tab:losv_catalogue} shows a part of the GC + UCD catalogue, which reports IDs, sky coordinates, and velocities, Table \ref{tab:foreground_stars} reports the foreground stars. The full version will be made available online.

In the future, these catalogue may be refined and additional parameters such as photometric measurements from literature work will be added where used in upcoming work. 

\begin{table}[]
    \centering
    \caption{Catalogue of LOS velocities.}
    \begin{tabular}{c c c c c} \hline\hline
    ID & RA & DEC & $v$ & $\Delta v$ \\
     & (J2000) & (J2000) & (km s$^{-1}$) & (km s$^{-1}$) \\ \hline
M49-GC1 & 187.42650 & 7.83050 & 758.54 & 1.47\\
M49-GC2 & 187.50287 & 7.96436 & 907.31 & 3.98\\
M49-GC3 & 187.41537 & 7.89244 & 1459.81 & 2.03\\
M49-GC4 & 187.45562 & 7.96481 & 987.13 & 5.86\\
M49-GC5 & 187.41321 & 7.78194 & 1140.61 & 2.77\\
M49-GC6 & 187.41137 & 7.91261 & 945.52 & 1.14\\
M49-GC7 & 187.40912 & 8.02886 & 1395.17 & 1.53\\
M49-GC8 & 187.43287 & 7.96578 & 1033.84 & 1.03\\
M49-GC9 & 187.41521 & 7.96747 & 1074.93 & 0.80\\
M49-GC10 & 187.37417 & 7.79336 & 1035.34 & 1.17\\
M49-GC11 & 187.37292 & 7.99631 & 1030.58 & 1.43\\
M49-UCD1 & 187.36921 & 7.97347 & 1184.98 & 1.08\\
M49-UCD2 & 187.34517 & 7.81692 & 1733.39 & 0.87\\ 
$\cdot\cdot\cdot$ & $\cdot\cdot\cdot$ & $\cdot\cdot\cdot$ & $\cdot\cdot\cdot$ & $\cdot\cdot\cdot$\\
\hline
    \end{tabular}
    \label{tab:losv_catalogue}
        \tablefoot{The full catalogue is available online at the CDS.}
\end{table}

\begin{table}[]
    \centering
    \caption{Catalogue of foreground stars.}
    \begin{tabular}{c c c c c} \hline\hline
    ID & RA & DEC & $v$ & $\Delta v$ \\
     & (J2000) & (J2000) & (km s$^{-1}$) & (km s$^{-1}$) \\ \hline
M87-Star2 & 187.87558 & 12.49297 & 32.47 & 0.36\\
M87-Star3 & 187.84958 & 12.50447 & 40.40 & 0.42\\
M87-Star5 & 187.62779 & 12.52408 & -32.07 & 0.15\\
M87-Star6 & 187.59267 & 12.53447 & -20.98 & 1.38\\
M87-Star7 & 187.68087 & 12.54678 & -30.29 & 0.62\\
M87-Star8 & 187.69287 & 12.55047 & 182.95 & 0.34\\
M87-Star9 & 187.70008 & 12.55497 & -5.86 & 0.75\\
M87-Star10 & 187.76308 & 12.55708 & 23.23 & 0.37\\
M87-Star11 & 187.85308 & 12.55778 & 27.97 & 0.84\\
M87-Star12 & 187.76308 & 12.56558 & 209.98 & 1.53\\
M87-Star13 & 187.81729 & 12.57789 & 76.76 & 0.38\\
M87-Star14 & 187.82779 & 12.62378 & 140.17 & 1.80\\
$\cdot\cdot\cdot$ & $\cdot\cdot\cdot$ & $\cdot\cdot\cdot$ & $\cdot\cdot\cdot$ & $\cdot\cdot\cdot$\\
\hline
    \end{tabular}
    \label{tab:foreground_stars}
        \tablefoot{The full catalogue is available online at the CDS.}
\end{table}

\end{appendix}
\end{document}